\documentclass[twoside,twocolumn,9pt]{article}
\usepackage{extsizes}
\usepackage[super,sort&compress,comma]{natbib}
\usepackage[version=3]{mhchem}
\usepackage[left=1.5cm, right=1.5cm, top=1.785cm, bottom=2.0cm]{geometry}
\usepackage{balance}
\usepackage{mathptmx}
\usepackage{sectsty}
\usepackage{graphicx}
\usepackage{lastpage}
\usepackage{amsmath}
\usepackage{bm}
\usepackage{amssymb}
\usepackage[format=plain,justification=justified,singlelinecheck=false,font={stretch=1.125,small,sf},labelfont=bf,labelsep=space]{caption}
\usepackage{float}
\usepackage{fancyhdr}
\usepackage{fnpos}
\usepackage[english]{babel}
\addto{\captionsenglish}{%
  
}
\usepackage{array}
\usepackage{droidsans}
\usepackage{charter}
\usepackage[T1]{fontenc}
\usepackage[usenames,dvipsnames]{xcolor}
\usepackage{setspace}
\usepackage[compact]{titlesec}
\usepackage{hyperref}

\usepackage{epstopdf}

\definecolor{cream}{RGB}{222,217,201}

\usepackage{mathtools}
\usepackage{caption}
\usepackage{subcaption}
\usepackage{ragged2e}

\usepackage{tikz}
\usetikzlibrary{arrows.meta}
\usetikzlibrary{decorations.pathreplacing,calligraphy}

\begin{document}

\pagestyle{fancy}
\thispagestyle{plain}
\fancypagestyle{plain}{

\renewcommand{\headrulewidth}{0pt}
}

\makeFNbottom
\makeatletter
\renewcommand\LARGE{\@setfontsize\LARGE{15pt}{17}}
\renewcommand\Large{\@setfontsize\Large{12pt}{14}}
\renewcommand\large{\@setfontsize\large{10pt}{12}}
\renewcommand\footnotesize{\@setfontsize\footnotesize{7pt}{10}}
\makeatother

\renewcommand{\thefootnote}{\fnsymbol{footnote}}
\renewcommand\footnoterule{\vspace*{1pt}%
\color{cream}\hrule width 3.5in height 0.4pt \color{black}\vspace*{5pt}}
\setcounter{secnumdepth}{5}
\makeatletter
\renewcommand\@biblabel[1]{#1}
\renewcommand\@makefntext[1]%
{\noindent\makebox[0pt][r]{\@thefnmark\,}#1}
\makeatother
\renewcommand{\figurename}{\small{Fig.}~}
\sectionfont{\sffamily\Large}
\subsectionfont{\normalsize}
\subsubsectionfont{\bf}
\setstretch{1.125} 
\setlength{\skip\footins}{0.8cm}
\setlength{\footnotesep}{0.25cm}
\setlength{\jot}{10pt}
\titlespacing*{\section}{0pt}{4pt}{4pt}
\titlespacing*{\subsection}{0pt}{15pt}{1pt}

\fancyfoot{}
\fancyfoot[RO]{\footnotesize{\sffamily{1--\pageref{LastPage} ~\textbar  \hspace{2pt}\thepage}}}
\fancyfoot[LE]{\footnotesize{\sffamily{\thepage~\textbar\hspace{3.45cm} 1--\pageref{LastPage}}}}
\fancyhead{}
\renewcommand{\headrulewidth}{0pt}
\renewcommand{\footrulewidth}{0pt}
\setlength{\arrayrulewidth}{1pt}
\setlength{\columnsep}{6.5mm}
\setlength\bibsep{1pt}

\makeatletter
\newlength{\figrulesep}
\setlength{\figrulesep}{0.5\textfloatsep}

\newcommand{\topfigrule}{\vspace*{-1pt}%
\noindent{\color{cream}\rule[-\figrulesep]{\columnwidth}{1.5pt}} }

\newcommand{\botfigrule}{\vspace*{-2pt}%
\noindent{\color{cream}\rule[\figrulesep]{\columnwidth}{1.5pt}} }

\newcommand{\dblfigrule}{\vspace*{-1pt}%
\noindent{\color{cream}\rule[-\figrulesep]{\textwidth}{1.5pt}} }

\makeatother

\twocolumn[
  \begin{@twocolumnfalse}
\vspace{1em}
\sffamily
\begin{tabular}{m{4.5cm} p{13.5cm} }

& \noindent\LARGE{\textbf{Self Consistent Field Theory of isotropic-nematic interfaces and disclinations in a semiflexible molecule nematic}} \\
\vspace{0.3cm} & \vspace{0.3cm} \\

 & \noindent\large{Longyu Qing and Jorge Vi\~nals} \\
  & \noindent\normalsize{A Self Consistent Field Theory description of equilibrium, but non uniform, configurations adopted by semi flexible liquid crystal molecules is presented.
  Two cases are considered, isotropic-nematic phase boundaries, and topological defects in the nematic phase (disclinations). Nematogens are modeled by worm-like chains, with microscopic interaction potential of the Maier-Saupe type, with an added isotropic excluded volume contribution. The thermodynamic fields obtained by numerical minimization of the free energy are the molecular density and the nematic tensor order parameter. Interfaces with both homeotropic and planar alignment are studied, as well as biaxiality and anisotropy around $\pm 1/2$ disclinations. The effects induced by fluid compressibility, interaction strength, and elastic anisotropy that follows from chain flexibility on both types of non uniform configurations are discussed. Defect core sizes decrease as the system becomes less compressible, eventually reaching a constant value in the incompressible limit. The core size is influenced by the nematic interaction strength $u_2$ and chain persistence length $l_p$, decreasing as the order increases in the nematic region through manipulation of $l_p$ and $u_2$. Additionally, when the far field nematic order is fixed, the core size increases with $l_p$.}

\end{tabular}
 \end{@twocolumnfalse} \vspace{0.6cm}

  ]

\renewcommand*\rmdefault{bch}\normalfont\upshape
\rmfamily
\section*{}
\vspace{-1cm}

\footnotetext{\textit{School of Physics and Astronomy, University of Minnesota,
Minneapolis, MN 55455, USA.}}


\section{Introduction}
\label{sec:introduction}

Self Consistent Field Theory (SCFT) is a well established technique for the study of the equilibrium properties of polymer and block copolymer melts by computing energy and entropy contributions to the partition function that arise from chain architecture, flexibility, and intermolecular interactions \cite{re:fredrickson02,re:fredrickson06}. Of particular interest here are systems comprising semi flexible chains that assemble into nematic and smectic ordered phases \cite{re:song09,jiang2010isotropic,deng2010wormlike,spencer2020nematic}. Our analysis is motivated by growing interest in orientationally ordered and often active fluids, including bio polymeric systems. In many cases, the molecular constituents are quite complex, and remain poorly characterized. Therefore the resulting elastic anisotropy of the medium, and its often unusual rheology remain largely unexplored. Self Consistent Field Theory offers a potential avenue for the exploration of elasticity and the structure of topological defects in the types of molecular assemblies that are the focus of current active and biological matter research. Our work presented below is our first step in that direction, and focuses on the structure of topological defects in a nematic phase comprising semi flexible molecular units. 

The free energy of elastic distortion from a uniform nematic phase is given, to lowest order, by three modes of deformation: splay, twist, and bend \cite{re:selinger16}. The starting point of many theoretical analyses of nematics, however, is the so called one constant approximation according to which splay, twist, and bend elastic constants are assumed to be equal. This renders the nematic elastically isotropic, a reasonably good approximation for small molecule, thermotropic liquid crystals. On the other hand, the response of the so called lyotropic liquid crystals (longer molecule nematogens in solution, where the isotropic to nematic transition is induced by concentration change) is quite different. Broad classes that are being studied at present include lyotropic chromonics \cite{re:zhou12,re:zhou17b}, and nematic micelles \cite{re:dietrich20}. In both cases, the twist elastic constant is as much as one order of magnitude smaller than bend and twist. Such a large contrast gives rise to unexpected phenomenology, including, for example, spontaneous chiral symmetry breaking under confinement \cite{re:jeong14,re:nayani15,re:davidson15,re:velez21,re:myers25}. Additional research on nematic response in systems with complex molecular units include, for example, actin networks \cite{re:zhang18}, the role of nematic order in cellular mechano adaptation \cite{re:cook23}, or growth of gliomas in the brain \cite{re:faisal24}. Theoretical tools are needed to describe potentially defected nematics exhibiting large elastic anisotropy, and with complex molecular architecture and rheology. Given the successes of Self Consistent Field Theory in the polymer field, we begin here by examining the effects of fluid compressibility and molecular flexibility on complex configurations of a nematic phase that involve phase boundaries and topological defects.

In the related case of worm like chains, it is known that both persistence ($l_p$) and chain lengths ($L$) determine the elastic anisotropy of the nematic phase \cite{varytimiadou2024elasticity}. Polymer field theory work \cite{ghosh2022semiflexible} shows that the bend elastic constant ($K_3$) is larger than the splay constant ($K_1$) for rigid chains($L\ll l_p$), while the splay constants is larger for flexible chains ($L \gg l_p$), indicating a crossover between splay and bend contrast as chain flexibility changes. For a wide range of flexibility, the twist elastic constant ($K_2$) is smaller than both splay and bend \cite{ghosh2022semiflexible, odijk1986elastic}. In particular, $K_{2}/K_{1,3}$ becomes small when $L \gg l_{p}$. Hence, inclusion of nematogen flexibility into the theory naturally leads to elastic anisotropy. In addition to considering chain flexibility, we introduce a molecular interaction potential of the Maier-Saupe type, supplemented by an isotropic excluded volume term. This allows a simultaneous study of an interaction inducing nematic order, and of fluid compressibility. As expected, we find that increasing density favors nematic order. The same effect follows by increasing the persistence length. For the case of an isotropic-nematic interface at coexistence, SCFT correctly predicts biaxiality in the interfacial region when the alignment is planar, and uniaxiality and a wider interface in the case of homeotropic alignment. We also show that the interfacial width increases with persistence length. Disclinations are also correctly reproduced within the theory, including a region of biaxiality near the core, crossing over to a uniaxial core. While the molecular density around the defect core depends strongly on system compressibility and the value of the Maier-Saupe interaction coefficient, the nematic order parameter is largely insensitive to the system compressibility. Finally, the angular dependence of the eigenvalues of the tensor order parameter found around $\pm 1/2$ disclinations is consistent with the elastic anisotropy induced by the flexible chains, in agreement with experiments \cite{re:zhou17}, and with calculations based on the singular potential method \cite{schimming2020anisotropic}.

\section{Self Consistent Field Theory}
\label{sec:SCFT}
In this section, and for completeness, we briefly summarize our implementation of the self-consistent field theory (SCFT) for semiflexible molecules with Maier-Saupe interaction. Similar derivations can be found in Ref. \cite{chen2016theory, spencer2020nematic}. The molecules are approximated by worm-like chains characterized by a contour length $L_c$, and a persistence length $l_p$. Each chain consists of $N_s$ segments or monomers. For a collection of $n$ worm-like chains confined within volume $V$ at temperature $T$, the configuration of the $i$-th chain is described by a space curve $\mathbf{r}_i(s)$, where $s$ ($0\leq s \leq 1$) is a normalized contour variable that denotes the location of a segment along the chain. The unit tangent vector $\mathbf{u}_i(s) = d\mathbf{r}_i /(L_c ds)$ gives the orientation of the segment $s$. The microscopic density of segment position and orientation is defined as
\begin{equation}
\hat{\phi}(\mathbf{r},\mathbf{u})= \frac{V}{n} \sum_{i=1}^{n} \int_{0}^{1} ds \; \delta(\mathbf{r}-\mathbf{r}_i(s))\delta(\mathbf{u}-\mathbf{u}_i(s))
\end{equation}
which is dimensionless and satisfies the normalization condition $\int d\mathbf{r} \int d \mathbf{u} \hat{\phi}(\mathbf{r},\mathbf{u}) = V$. For a spatially homogeneous system, the average density $\phi(\mathbf{r},\mathbf{u}) =  \langle \hat{\phi}(\mathbf{r},\mathbf{u}) \rangle$ is independent of $\mathbf{r}$, where $\langle . \rangle$ denotes a thermal average. Consequently, the normalization condition then implies that the average density, when integrated over all orientations, satisfies $\phi(\mathbf{r}) = \int d\mathbf{u}\phi(\mathbf{r},\mathbf{u})=1$. In contrast, for inhomogeneous systems, the local average density at $\mathbf{r}$ may take values that are greater or less than one.

The energy of the system consists of a bending energy of individual nematoges, monomer-monomer interactions and monomer-solvent interactions. The solvent is treated implicitly, and the monomer-monomer interactions are assumed to follow the Maier-Saupe model. The Hamiltonian of the system is given by
\begin{equation}
\begin{aligned}
\beta H &=\frac{l_p}{2L_c} \sum_{i=1}^n \int_{0}^{1} ds \left| \frac{d\mathbf{u}_i(s)}{ds} \right|^2 + \\
&\frac{1}{2}\frac{n^2}{V^2}\int d\mathbf{r}\int d\mathbf{r}' \int d\mathbf{u}\int d\mathbf{u}' \ \hat{\phi}(\mathbf{r},\mathbf{u}) V(\mathbf{r}, \mathbf{r}';\mathbf{u}, \mathbf{u}')\hat{\phi}(\mathbf{r}',\mathbf{u}')
\end{aligned}
\end{equation}
where $\beta = 1/k_{B}T$, and the kernel function $V(\mathbf{r}, \mathbf{r}';\mathbf{u}, \mathbf{u}')=\delta(\mathbf{r}-\mathbf{r}')\{ u_0- u_2 [(\mathbf{u} \cdot \mathbf{u}')^2-\frac{1}{3}]\}$ in the present study. The excluded volume parameter $u_0$ quantifies the strength of the isotropic interaction, and $u_2$ quantifies the strength of the anisotropic Maier-Saupe (MS) interaction. The resulting partition function, after a Hubbard-Stratonovich transformation, is
\begin{equation}
Z \propto \int D \phi \int D w \exp(-\beta F[\phi, w]) 
\end{equation}
with a free energy functional given by,
\begin{equation}
\begin{aligned}
\beta F[\phi, w] &= -  \frac{n}{V}\int d \mathbf{r} \int d \mathbf{u} \ w(\mathbf{r}, \mathbf{u}) \phi(\mathbf{r}, \mathbf{u})   - n\ln Z_{1} [w] + n\ln(\frac{n}{V}) +  nF_0\\
&+ \frac{1}{2}\frac{n^2}{V^2}\int d \mathbf{r} \int d \mathbf{r}' \int d \mathbf{u} \int d \mathbf{u}' \phi(\mathbf{r}, \mathbf{u}) V(\mathbf{r}, \mathbf{r}';\mathbf{u}, \mathbf{u}') \phi(\mathbf{r}', \mathbf{u}') \label{eq:FreeEnergy}
\end{aligned}
\end{equation}
where $F_0$ is a constant that is independent of $n$ and $V$, and $Z_{1}[w]$ is the normalized single chain partition function that is a functional of any specified external field $w(\mathbf{r}, \mathbf{u})$. Note that this functional, as defined, is non local. Hence the free energy functional of Eq. \eqref{eq:FreeEnergy} cannot be written as an integral, over the entire system, of a free energy density. Next, a saddle point approximation assumes that the extremal configurations $[\phi^*, w^*]$ dominate the functional integral defining the partition function, and are defined by,
\begin{equation}
\left.\frac{\delta (\beta F)}{\delta \phi} \right|_{\phi=\phi^*, w=w^*}=0 , \quad \quad \left. \frac{\delta (\beta F)}{\delta w} \right|_{\phi=\phi^*, w=w^*}=0 . 
\label{eq:sp}
\end{equation}
Thus the free energy is approximated by $F[\phi^*, w^*]$. The saddle point approximation yields the relations
\begin{equation}
\begin{aligned}
w(\mathbf{r}, \mathbf{u}) &= \frac{n}{V}\int d \mathbf{r}' \int d \mathbf{u}' V(\mathbf{r}, \mathbf{r}';\mathbf{u}, \mathbf{u}') \phi(\mathbf{r}', \mathbf{u}')\\
&=\frac{n}{V} \int d \mathbf{u}'  \left[ u_0- u_2 [(\mathbf{u} \cdot \mathbf{u}')^2-\frac{1}{3}] \right]\phi(\mathbf{r}', \mathbf{u}')\\
\phi (\mathbf{r}, \mathbf{u}) &=  -V \frac{\delta \ln Z_{1}[w]}{\delta w} \label{eq:sc}
\end{aligned} 
\end{equation}
for the two thermodynamically independent fields $w(\mathbf{r}, \mathbf{u})$ and $\phi (\mathbf{r}, \mathbf{u})$. Notably, the second equation involves solving for a single chain in external field $w(\mathbf{r}, \mathbf{u})$
\begin{equation}
\begin{aligned}
\phi (\mathbf{r}, \mathbf{u}) &= \int_0^1 ds  \phi (\mathbf{r}, \mathbf{u}, s) \\
&= \frac{1}{4 \pi Z_{1}[w]}\int_0^1 ds q(\mathbf{r}, -\mathbf{u}, 1-s;[w])q(\mathbf{r}, \mathbf{u}, s;[w])
\end{aligned}
\end{equation}
where the propagator $q$ is a functional of $w(\mathbf{r}, \mathbf{u})$, satisfying the modified diffusion equation (MDE)
\begin{equation}
\frac{\partial q( \mathbf{r},  \mathbf{u}, s;[w])}{\partial s} = \left( \frac{L_c}{2l_p} \nabla^2_{\mathbf{u}} - L_c \mathbf{u} \cdot \nabla_{\mathbf{r}} - w(\mathbf{r}, \mathbf{u})\right)q ,
\end{equation}
with initial condition $q(\mathbf{r},  \mathbf{u}, s=0;[w])=1$. The normalized single chain partition function is given by
\begin{equation}
Z_{1}[w] = \frac{1}{4 \pi V} \int d \mathbf{r} \int d \mathbf{u} q( \mathbf{r},  \mathbf{u}, s=1;[w]) 
\end{equation}
The equilibrium solutions are determined self consistently according to Eqs. \ref{eq:sc}. It means that the interaction between molecules for a configuration $\phi(\mathbf{r}, \mathbf{u})$ creates an effective field $w(\mathbf{r}, \mathbf{u})$. This generated field in turn self consistently determines the configuration $\phi(\mathbf{r}, \mathbf{u})$. The equilibrium states are obtained when both self consistency conditions are simultaneously satisfied. 

The nematic order parameter tensor is obtained from $\phi (\mathbf{r}, \mathbf{u})$ through
\begin{equation}
\mathbf{Q}(\mathbf{r}) = \frac{\int d \mathbf{u} \ \phi (\mathbf{r}, \mathbf{u}) \left( \mathbf{u} \otimes \mathbf{u} - \frac{1}{3} \mathbf{I} \right)}{\int d \mathbf{u} \ \phi (\mathbf{r}, \mathbf{u})} \label{eq:tensor_def}
\end{equation}
where $\mathbf{I}$ is the rank 3 identity tensor. The denominator is needed for normalization in the case of inhomogeneous states where the local density $\phi(\mathbf{r}) $ is not uniform.

In order to reduce the computational complexity, we expand all $\mathbf{u}$ dependent quantities in terms of real spherical harmonics\cite{spencer2020nematic, homeier1996some}. The real spherical harmonic expansion of an arbitrary function $f(\mathbf{r}, \mathbf{u})$ is written as 
\begin{equation}
f(\mathbf{r}, \mathbf{u}) = \sum_{l , m} f_l^m(\mathbf{r}) \tilde{Y}_l^m(\mathbf{u})
\end{equation}
where $f$ can be $q(\mathbf{r}, \mathbf{u}, s)$, $w(\mathbf{r}, \mathbf{u})$ and $\phi(\mathbf{r}, \mathbf{u})$. The MDE can then be written in terms of the spherical harmonic expansion as,
\begin{equation}
\frac{\partial q_l^m}{\partial s} = -\frac{L_c}{2 l_p} l(l+1) q_l^m - L_c \sqrt{\frac{4 \pi}{3}} G_{1l l_2}^{m_\alpha m m_2} \frac{\partial q_{l_2}^{m_2}}{\partial x_{\alpha}} - G_{l l_1 l_2}^{m m_1 m_2} w_{l_1}^{m_1} q_{l_2}^{m_2}
\end{equation}
where the real Gaunt coefficients are defined as $G_{l_1 l_2 l_3}^{m_1 m_2 m_3} = \int d \mathbf{u} \tilde{Y}_{l_1}^{m_1}(\mathbf{u})\tilde{Y}_{l_2}^{m_2}(\mathbf{u})\tilde{Y}_{l_3}^{m_3}(\mathbf{u})$ \cite{homeier1996some}, the integral of products of three real spherical harmonics over $\mathbf{u}$. The resulting convection diffusion equation for the coefficients is solved with a Lax-Wendroff method \cite{daoulas2005self}, with a step size in contour length $\Delta s$. The coordinate space is divided into evenly spaced lattice points, $(x_i, y_j, z_k)$. A field $f(\mathbf{r}, \mathbf{u})$ is represented by its expansion coefficients defined on lattice points, $f_l^m(x_i, y_j, z_k)$. In the computations below, we restrict the expansion to order $l=4$. The first equation in the system (\ref{eq:sc}) leads to linear equations following the expansion \cite{spencer2020nematic}
\begin{equation}
w_0^0(\mathbf{r}) = 4 \pi u_0 \frac{n}{V} \phi_0^0(\mathbf{r}) \quad\quad
w_2^m(\mathbf{r}) = -\frac{4 \pi u_2 }{5} \frac{n}{V} \phi_2^m(\mathbf{r}) 
\label{eq:scws}
\end{equation}
where all other components $l\neq0, 2$ are zero. Therefore, the expansion in real spherical harmonics greatly reduces the number of degrees of freedom of the theory. The expansion also allows us to write the order parameter tensor in terms of the coefficients $\phi_l^m$, following the definition in Eq.(\ref{eq:tensor_def})
\begin{equation}
\mathbf{Q}=\frac{1}{3 \sqrt{5} \phi_0^0}
\begin{pmatrix}
-\phi_2^0 +\sqrt{3}\phi_2^2 & \sqrt{3}\phi_2^{-2} & \sqrt{3}\phi_2^1 \\
\sqrt{3}\phi_2^{-2} & -\phi_2^0 - \sqrt{3}\phi_2^2 & \sqrt{3}\phi_2^{-1} \\
\sqrt{3}\phi_2^1 & \sqrt{3}\phi_2^{-1} & 2 \phi_2^0
\end{pmatrix}   \label{eq:q_spherical}
\end{equation}
where only $l=0,  2$ terms are nonzero due to the definition of $\mathbf{Q}$. The five degrees of freedom for $l=2$ are sufficient for $\mathbf{Q}$ to describe general biaxial order. As is done often, we parametrize the tensor order parameter as
\begin{equation}
\mathbf{Q}=S (\hat{\mathbf{n}} \otimes \hat{\mathbf{n}}-\frac{1}{3}\mathbf{I}) +P (\hat{\mathbf{m}} \otimes \hat{\mathbf{m}}-\hat{\mathbf{l}} \otimes \hat{\mathbf{l}})   \label{eq:qpara}
\end{equation}
where $\hat{\mathbf{n}}$, $\hat{\mathbf{m}}$ and $\hat{\mathbf{l}}$ are an orthonormal triad of eigenvectors of $\mathbf{Q}$. The director $\hat{\mathbf{n}}$ is the eigenvector with the largest eigenvalue. The constant $S$ is the uniaxial order parameter, and $P$ the biaxial order parameter. Their relationship with the eigenvalues of $\mathbf{Q}$ is, $(\lambda_n,\lambda_m,\lambda_l)=(\frac{2}{3}S,-\frac{1}{3}S+P,-\frac{1}{3}S-P)$.

There are only a limited number of cases in which the propagator can be obtained analytically, so a numerical solution to the MDE is usually necessary. In this work, in order to find approximate saddle point solutions (Eq. \eqref{eq:sp}), an iterative method is used with the following steps. Step 1: Define an initial guess for the input fields $w_{l(in)}^m(x_i, y_j, z_k)$. Step 2: Solve the MDE numerically to find  the propagator $q_l^m(x_i, y_j, z_k, s; [w_l^m])$, and compute $Z_{1}[w]$, $\phi_l^m(x_i, y_j, z_k)$, and other thermodynamic quantities accordingly. Step 3: Compute the output fields $w_{l(out)}^m(x_i, y_j, z_k)$ according to the first equation in (\ref{eq:sc}). Step 4: Define the absolute difference between the input and the output fields to be the error. When the maximum error among all the expansion components and space points is smaller than a given tolerance, i.e., $max(|w_{l(out)}^m(x_i, y_j, z_k) -w_{l(in)}^m(x_i, y_j, z_k)|) < \epsilon$, the iteration is ended. Otherwise, the input field is updated by using the Picard iteration method \cite{jiang2010isotropic, deng2010wormlike}, and going back to step 2. The current work obtains numerical solutions using Python\cite{qing_2025_14927168}.

\section{Uniform states and isotropic-nematic phase transition}
\label{sec:uniform}

We first briefly review the equilibrium phase behavior provided by the SCFT. We have used $\Delta s = 1/100$ in all our calculations. Unless otherwise specified, we set $u_0=10$, $L_c=l_p = 1$, and $n/V=1$. The normalization condition requires $\phi(\mathbf{r}) = 1$ and hence $\phi_0^0(\mathbf{r})=1 / \sqrt{4 \pi}$. Without loss of generality, we consider that the phase transition is from the isotropic phase to a uniaxial nematic phase along the $ \hat{\mathbf{z}}$ direction, so that $\phi_2^m$ is non vanishing only for $m=0$ in the nematic phase. Hence Eq.(\ref{eq:q_spherical}) gives the uniaxial order parameter $S=\sqrt{\frac{4 \pi}{5}}\phi_2^0$. 

\begin{figure*}[t]
\captionsetup[subfigure]{justification=Centering}
\hfill
\begin{subfigure}{0.42\textwidth}
 \includegraphics[width=\textwidth]{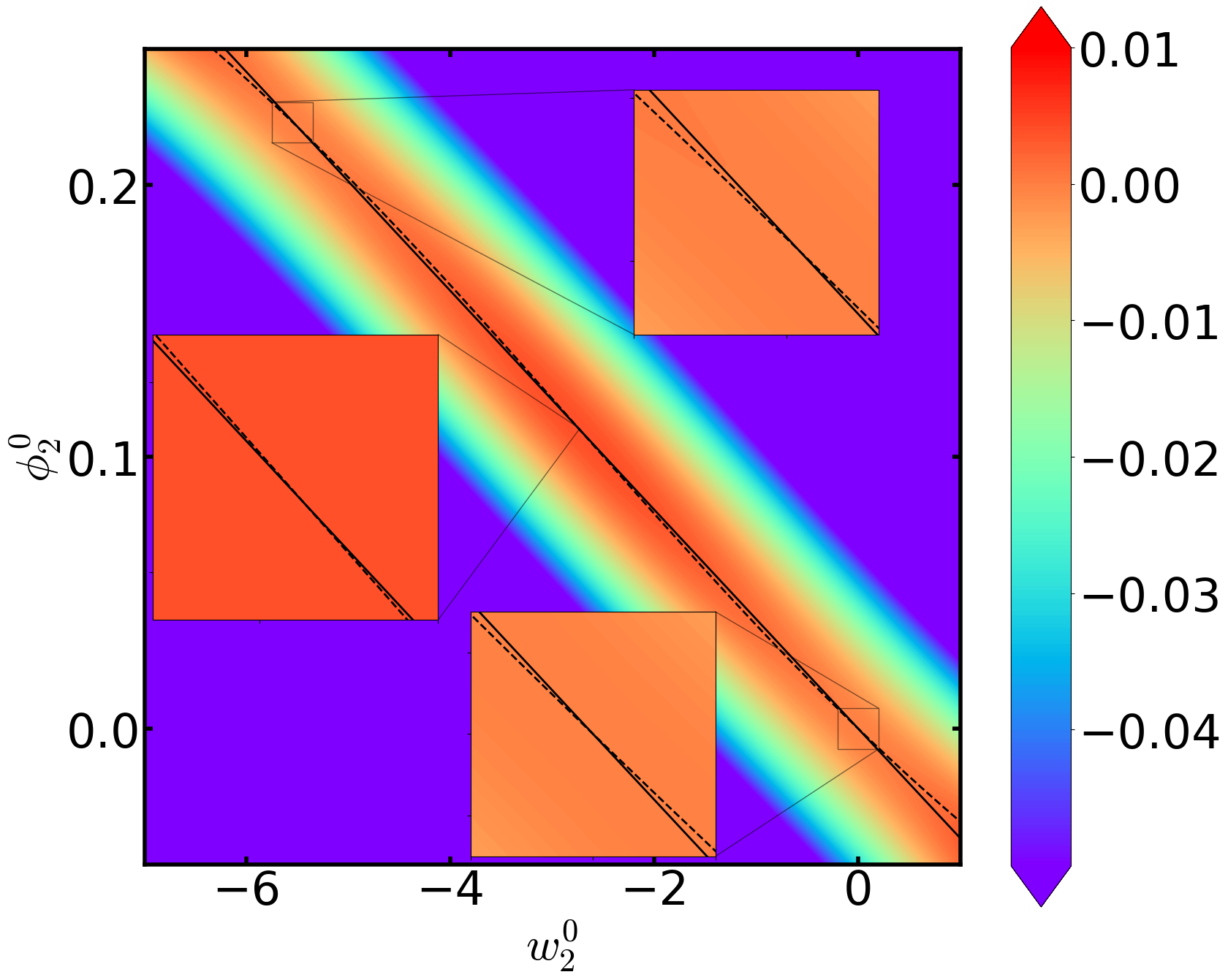}
    \caption{}
    \label{fig:free-energy-w-phi}
\end{subfigure}
\hfill
\begin{subfigure}{0.36\textwidth}
 \includegraphics[width=\textwidth]{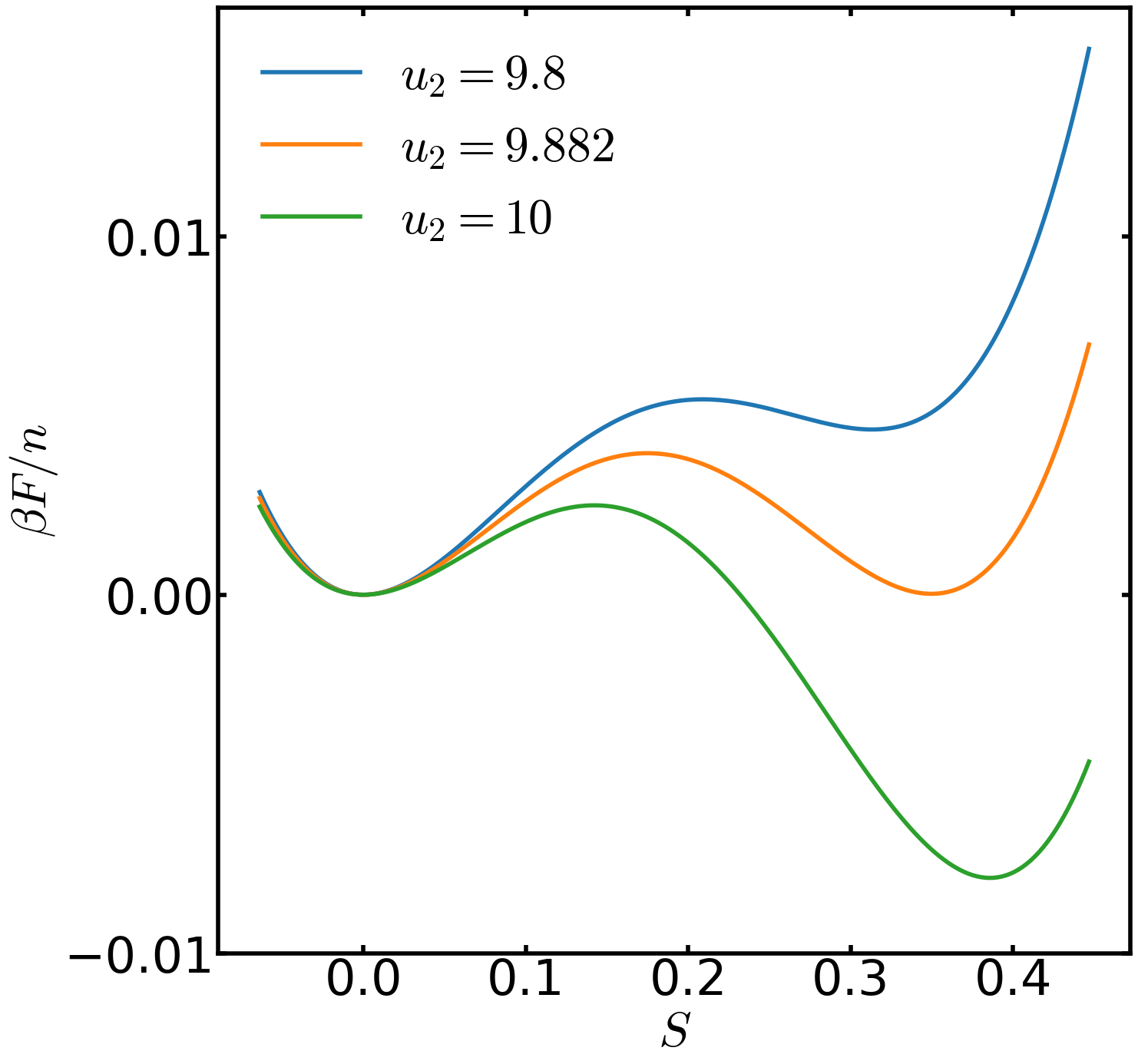}
    \caption{}
    \label{fig:free-energy-S}
\end{subfigure}
 \hfill

 \hfill
\begin{subfigure}{0.4\textwidth}
 \includegraphics[width=\textwidth]{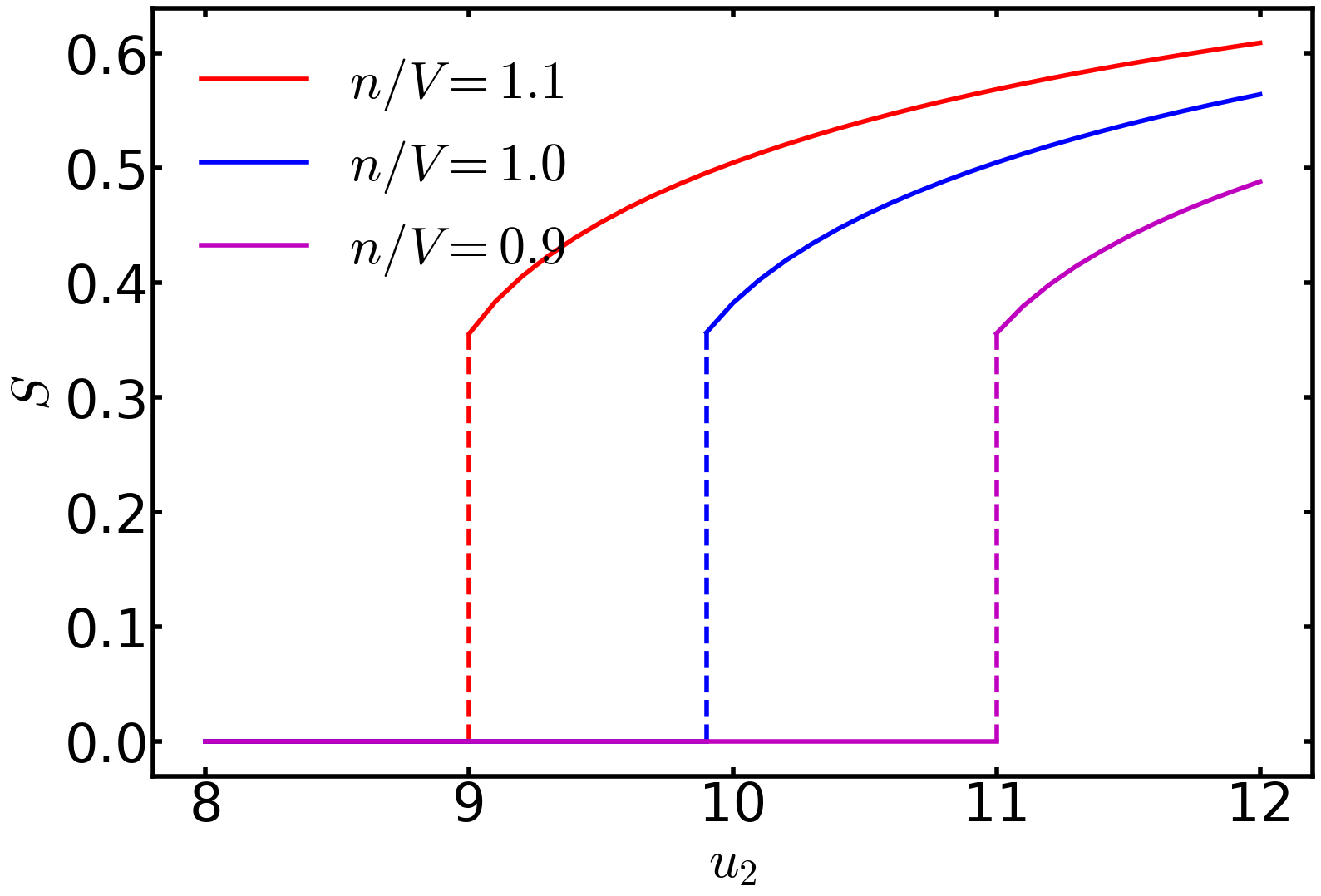}
    \caption{}
    \label{fig:S-u2-nV}
\end{subfigure}
\hfill
\begin{subfigure}{0.4\textwidth}
 \includegraphics[width=\textwidth]{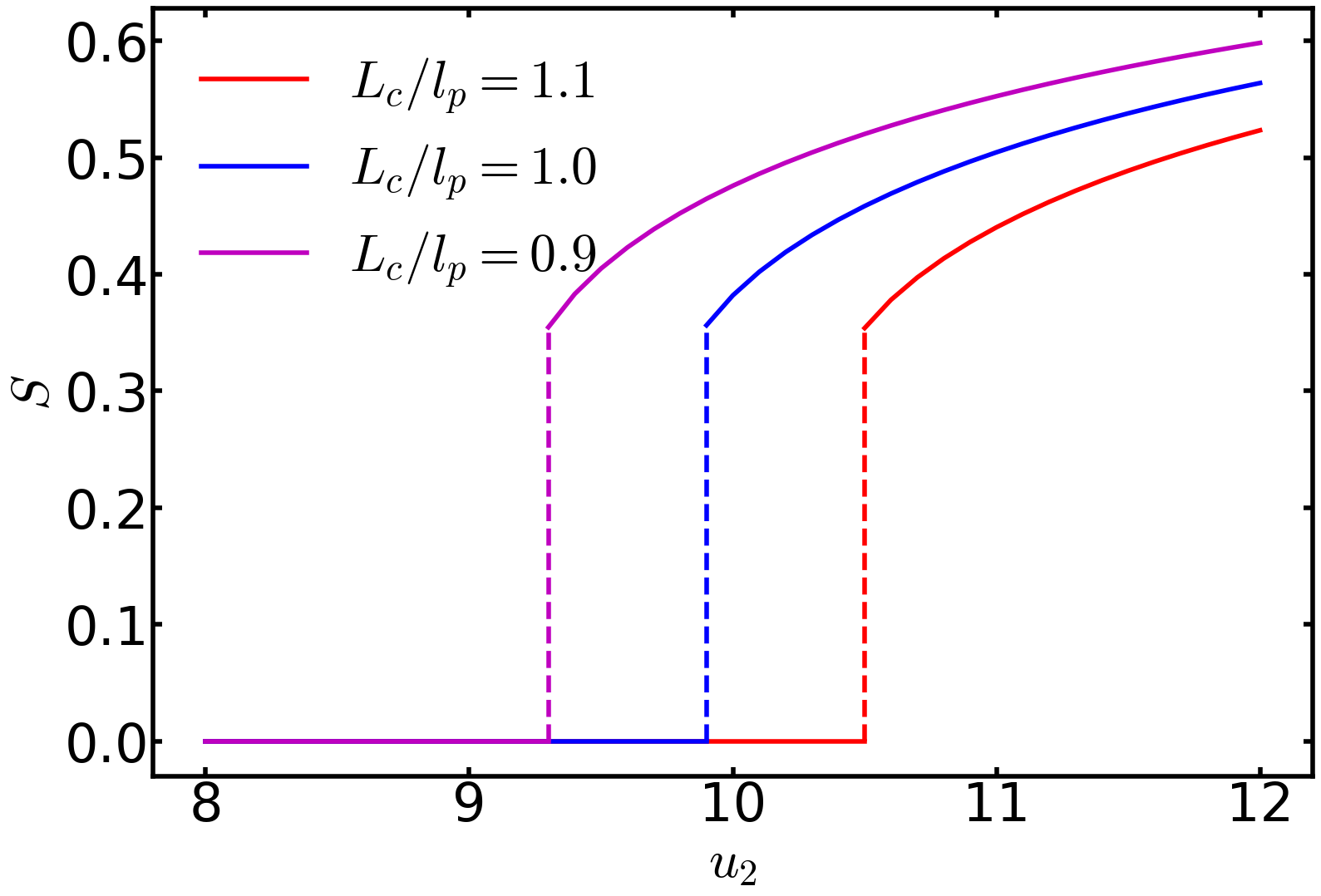}
    \caption{}
    \label{fig:S-u2-lp}
\end{subfigure}
 \hfill
\caption{(a) and (b)Free energy per chain of bulk state, where $F_0$ is chosen to make the free energy of isotropic phase zero. (a) shows the free energy per chain as function of $\phi_2^0$ and $w_2^0$ at the phase transition $u_2=9.882$, where the black solid line and dashed line correspond to the conditions $\frac{\delta (\beta F)}{\delta \phi_2^0}=0$ and $\frac{\delta (\beta F)}{\delta w_2^0}=0$, respectively. The three inset plots provide the enlarged view near the intersections. (b) shows the free energy per chain along the black solid line for different values of $u_2$. (c) and (d) Equilibrium uniaxial order $S$ as a function of $u_2$ for different chain number densities and flexibilities}  
\label{fig:bulk}
\end{figure*}

As described in Sec. \ref{sec:SCFT}, an iterative method is generally necessary to compute SCFT solutions from an appropriate initial guess. For a uniform configuration, and to the order of approximation in spherical harmonics that we are using, there is a more direct method. There are only two variables that need to be determined, $w_2^0$ and $\phi_2^0$. The free energy per chain as a function of  $w_2^0$ and $\phi_2^0$ is then given by,
\begin{equation}
\frac{\beta F[\phi_2^0, w_2^0]}{n} = F_0
 -  \ln Z_{1} [w_2^0] -w_2^0 \phi_2^0 - \frac{1}{2}\frac{n}{V}\frac{4 \pi u_2}{5} (\phi_2^0)^2
\end{equation}
where all constants are absorbed into $F_0$. The resulting free energy is shown in Fig.\ref{fig:bulk}(a) for $u_2=9.882$, the isotropic-nematic phase transition point. The black solid line illustrates the stationary condition $\frac{\delta (\beta F)}{\delta \phi_2^0}=0$, whereas the dashed line corresponds to $\frac{\delta (\beta F)}{\delta w_2^0}=0$. The saddle points are at the intersections of the solid and dashed lines. The equilibrium free energy per chain along the solid line ($w_2^0= -\frac{4 \pi u_2 }{5} \frac{n}{V} \phi_2^0$) displays a characteristic double well shape, and it is shown in Fig. \ref{fig:bulk}(b) for different values of $u_2$, and as a function of the uniaxial order parameter $S$. The saddle points corresponds to the minima of this function. The minimum at $S=0$ corresponds to the isotropic phase, and the minimum at $S\neq0$ corresponds to the nematic phase. The figure illustrates how, as $u_2$ increases, the global minimum changes from the isotropic phase to the nematic phase, indicating a first order phase transition. The equilibrium uniaxial order parameter $S$ as a function of MS interaction coefficient $u_2$ can be obtained from the minimum the free energy. The value of $u_2$ at the phase transition depends on combinations of parameters: Chain number density $n/V$ and ratio $L_c/l_p$, while it is independent of $u_0$. Fig.\ref{fig:bulk}(c)(d) show $S$ as a function of $u_2$ for different values of  $n/V$ and $L_c/l_p$. As the chain number density increases, the phase transition occurs at a lower value of $u_2$. Note that $u_2$ appears as a product with $n/V$ in Eq.\ref{eq:sc}, so the three curves in Fig.\ref{fig:bulk}(c) will be identical if $\tilde{u_2}=u_2n/V$ is used as the horizontal axis. As the chains become stiffer, Fig.\ref{fig:bulk}(d) shows that the phase transition occurs at a lower value of $u_2$, which is consistent with the result in Ref. \cite{spencer2020nematic}.

\section{Isotropic-Nematic Interface} \label{sec:interface}

\begin{figure*}[t]
\captionsetup[subfigure]{justification=Centering}
\hfill
\begin{subfigure}{0.4\textwidth}
 \includegraphics[width=\textwidth]{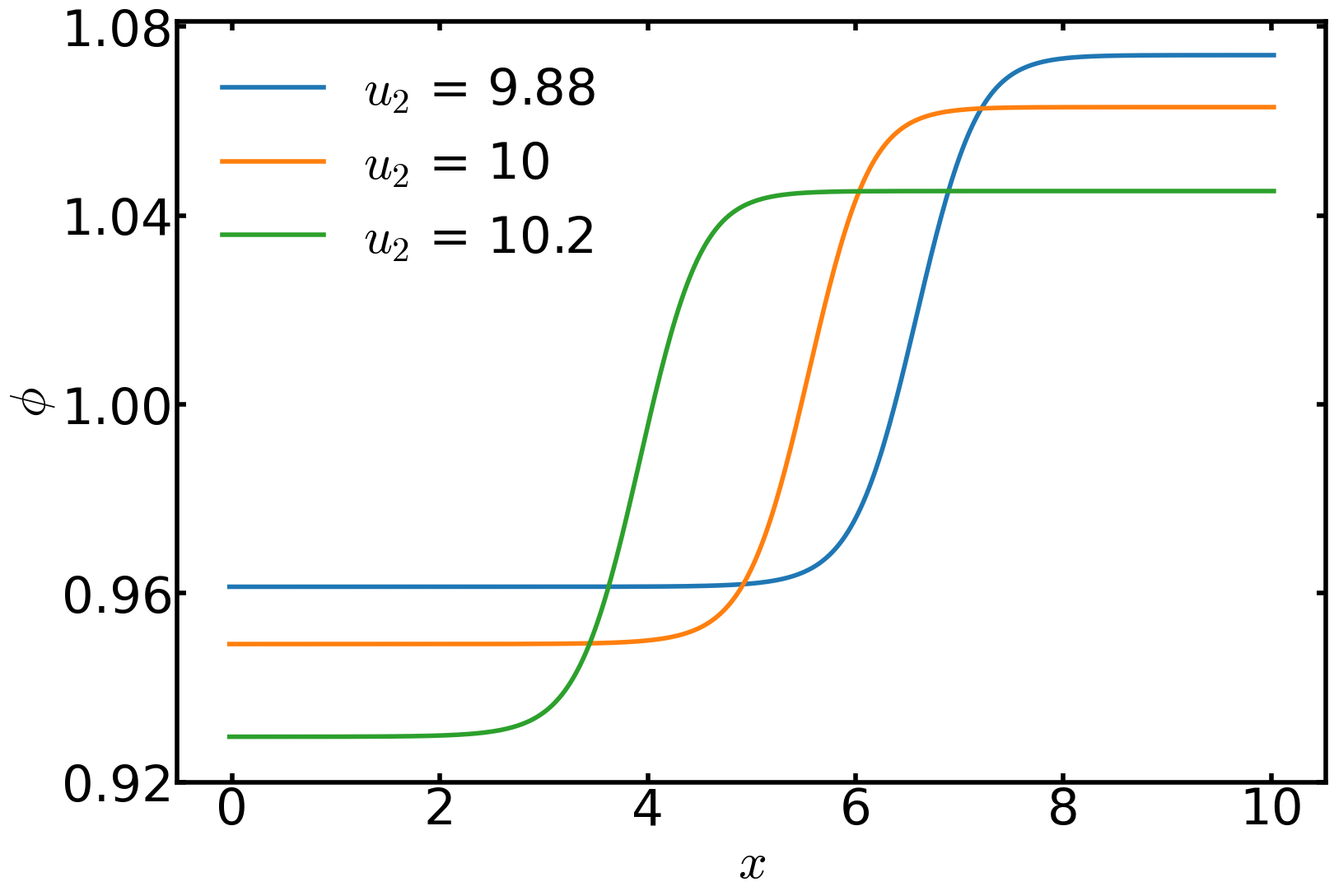}
    \caption{}
    \label{fig:u2-phi}
\end{subfigure}
\hfill
\begin{subfigure}{0.4\textwidth}
 \includegraphics[width=\textwidth]{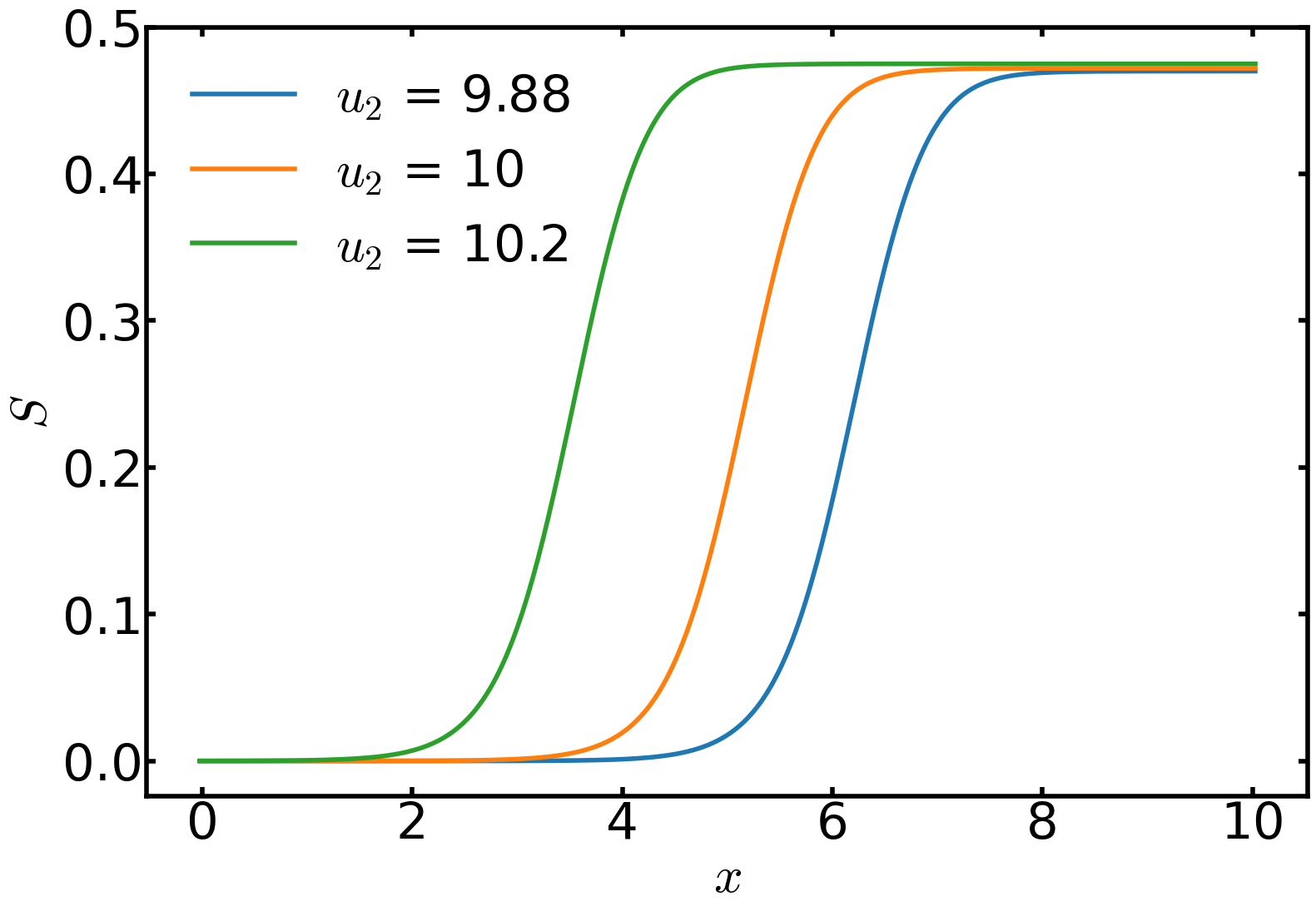}
    \caption{}
    \label{fig:u2-S}
\end{subfigure}
 \hfill

 \hfill
\begin{subfigure}{0.4\textwidth}
 \includegraphics[width=\textwidth]{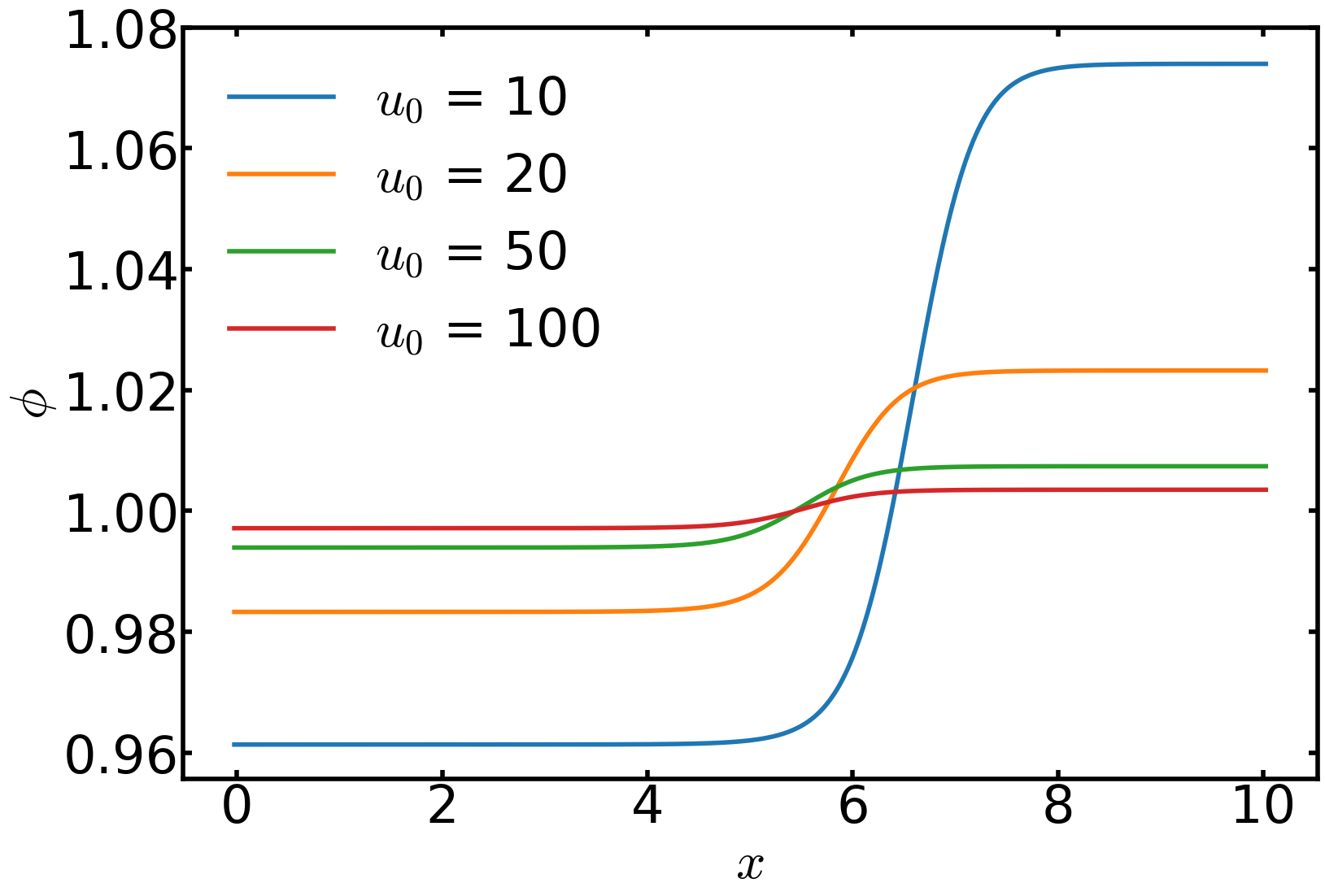}
    \caption{}
    \label{fig:u0-phi}
\end{subfigure}
\hfill
\begin{subfigure}{0.4\textwidth}
 \includegraphics[width=\textwidth]{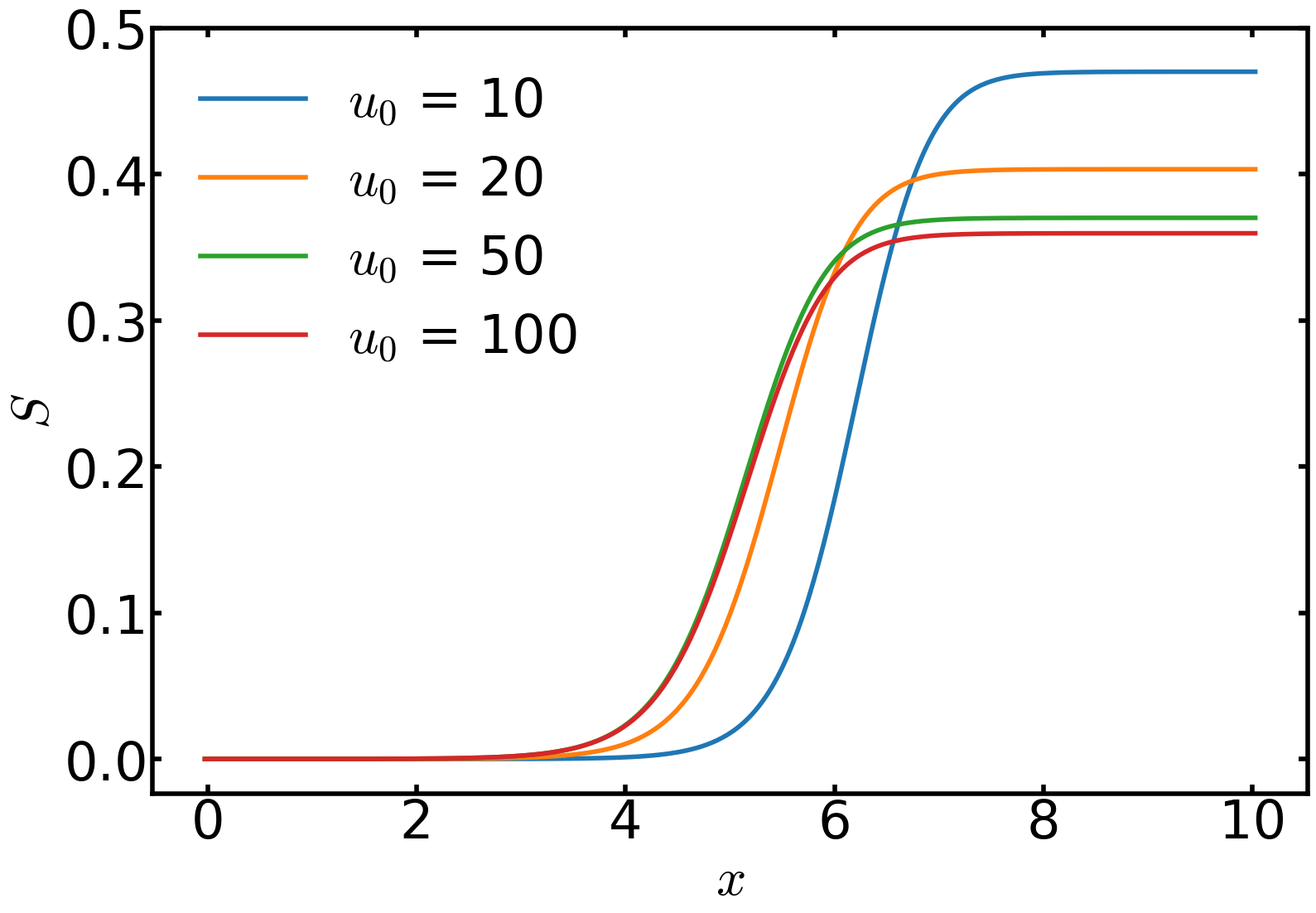}
    \caption{}
    \label{fig:u0-S}
\end{subfigure}
 \hfill

\caption{Isotropic-Nematic interface for different values of $u_2$ and $u_0$ in (a)(c) density profiles and (b)(d) uniaxial order profiles}  
\label{fig:u0-u2}
\end{figure*}

We study an inhomogeneous configuration with coexisting isotropic and nematic regions, separated by a planar interface. Unlike the case of Sec. \ref{sec:uniform} in which the density $\phi(\mathbf{r})$ is fixed and uniform in space, the density here is a function of position, and changes across the isotropic-nematic interface. At coexistence, in addition to uniform temperature, the chemical potential  $\mu = \left( \partial F/\partial n \right)|_{V}$ and pressure $p = - \left( \partial F/\partial V \right)|_{n}$ need to be the same in both phases. In addition, there is an equilibrium condition associated with free energy minimization with respect to the nematic order parameter. Far from the interface, the nematic phase is assumed uniaxial. Since the order parameter $S$ is unconstrainable, at coexistence one simply has $\left( \partial F/\partial S \right)|_{n, V} = 0$ in both bulk regions. The equilibrium configuration that contains an interface is obtained as follows: An initial configuration is set up in which $\mathbf{Q}$ and $\phi$ are fields varying only in the $\hat{\mathbf{x}}$ direction. To model an isotropic-nematic interface, a step function is introduced for $S$, where $S=0$ on the left side, representing the isotropic phase, and $S\neq0$ on the right side, representing the nematic phase. The initial value of $\phi_{0}^{0}=1/\sqrt{4 \pi}$ and $P=0$ are chosen for a uniform density configuration with zero biaxiality. For an arbitrary director direction $\hat{\mathbf{n}}$ in the nematic region, Eqs. (\ref{eq:scws})(\ref{eq:q_spherical})(\ref{eq:qpara}) are used to compute the initial values of $w_l^m$. Equations (\ref{eq:sc}) are iterated over a system of length $L_x=10L_c$ with Neumann boundary conditions applied on both ends. The domain $[0, 10]$ is uniformly divided into 400 intervals. The iteration process is terminated when the maximum errors in $w_l^m$ falls below $\epsilon=10^{-4}$. The iteration converges slowly if the initial interface is positioned arbitrarily. To accelerate convergence, the interface location is manually adjusted until the error rapidly decreases to $10^{-4}$. Although the initial condition assumes a uniform density configuration, a density gap between the isotropic and nematic regions naturally develops during the iteration process for a finite value of $u_0$. Furthermore, our results reveal the emergence of nonzero biaxiality across the interface under planar alignment, even though the initial condition assumes $P=0$. The nonzero biaxiality is not exclusive to planar alignment but is observed for any alignment where the director is not perpendicular to the interface \cite{jiang2010isotropic}.

\begin{figure*}[t]
\captionsetup[subfigure]{justification=Centering}
\hfill
\begin{subfigure}{0.4\textwidth}
 \includegraphics[width=\textwidth]{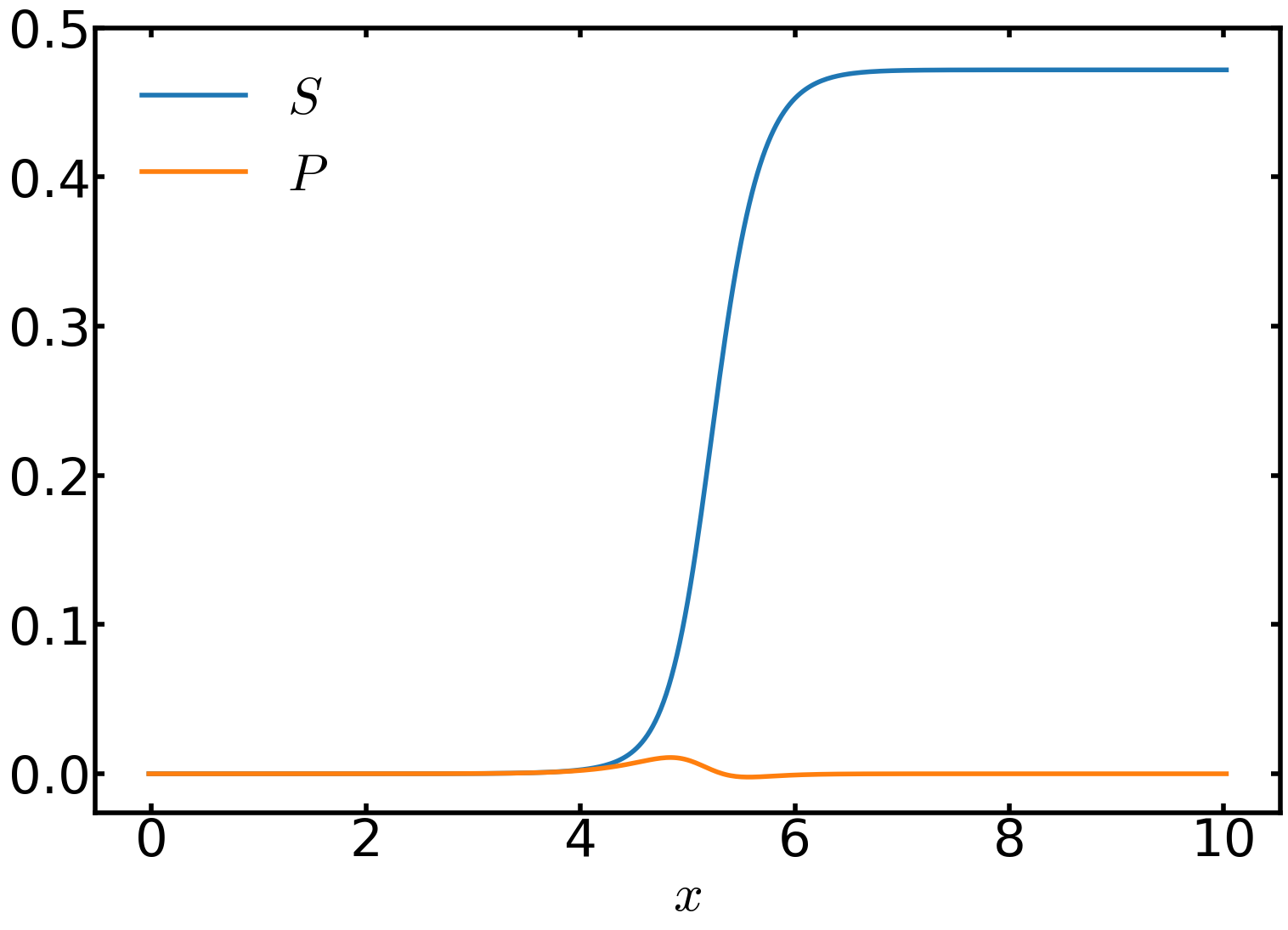}
    \caption{}
    \label{fig:free-energy-w-phi}
\end{subfigure}
\hfill
\begin{subfigure}{0.4\textwidth}
 \includegraphics[width=\textwidth]{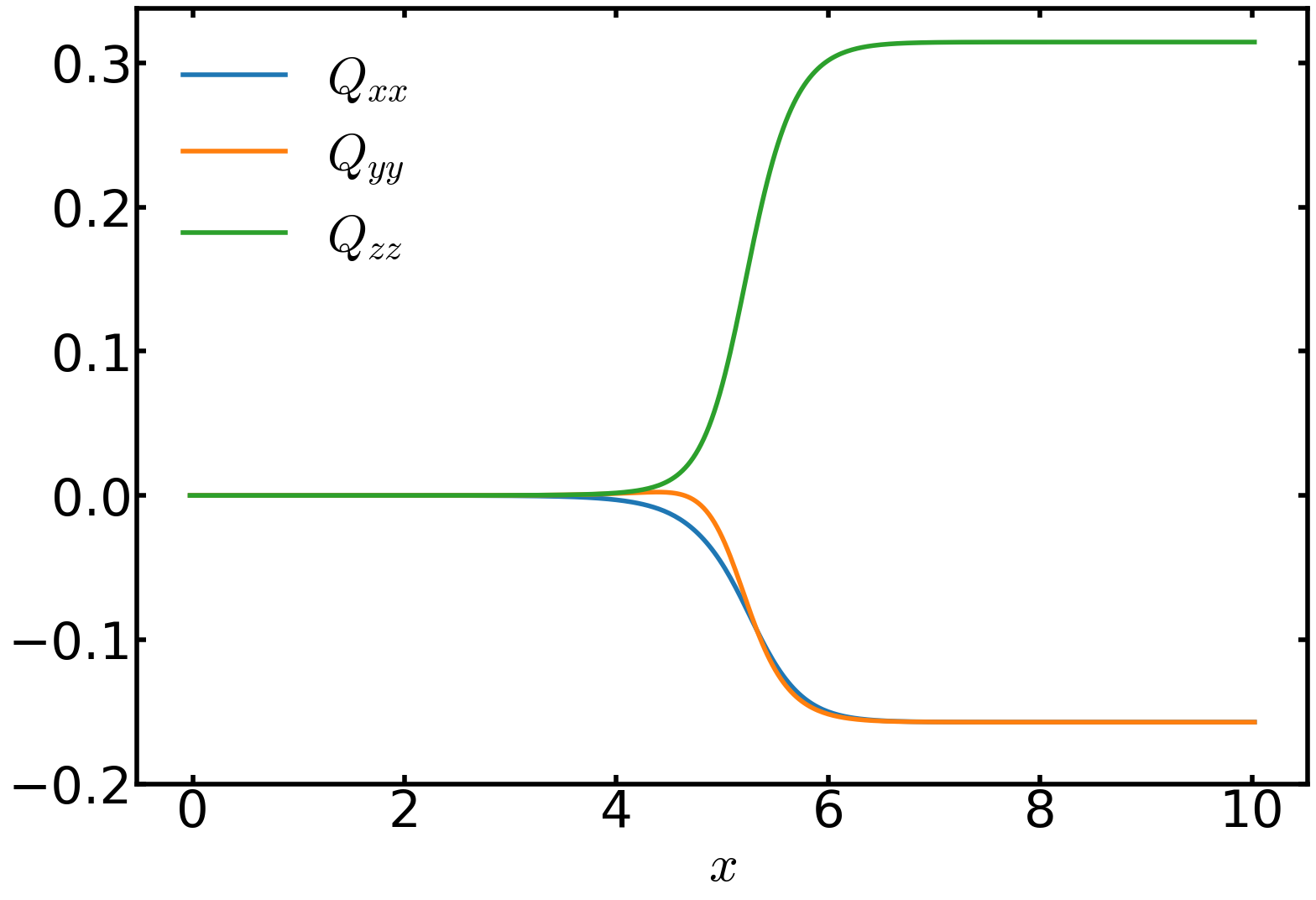}
    \caption{}
    \label{fig:free-energy-S}
\end{subfigure}
 \hfill
\caption{Interface for planar alignment. (a) shows the uniaxial order S and biaxial order P across the interface. (b) shows the three diagonal components of the tensor order parameter as functions of $x$ }
\label{fig:planar}
\end{figure*}

\begin{figure*}[t]
\captionsetup[subfigure]{justification=Centering}
 \hfill
\begin{subfigure}{0.4\textwidth}
 \includegraphics[width=\textwidth]{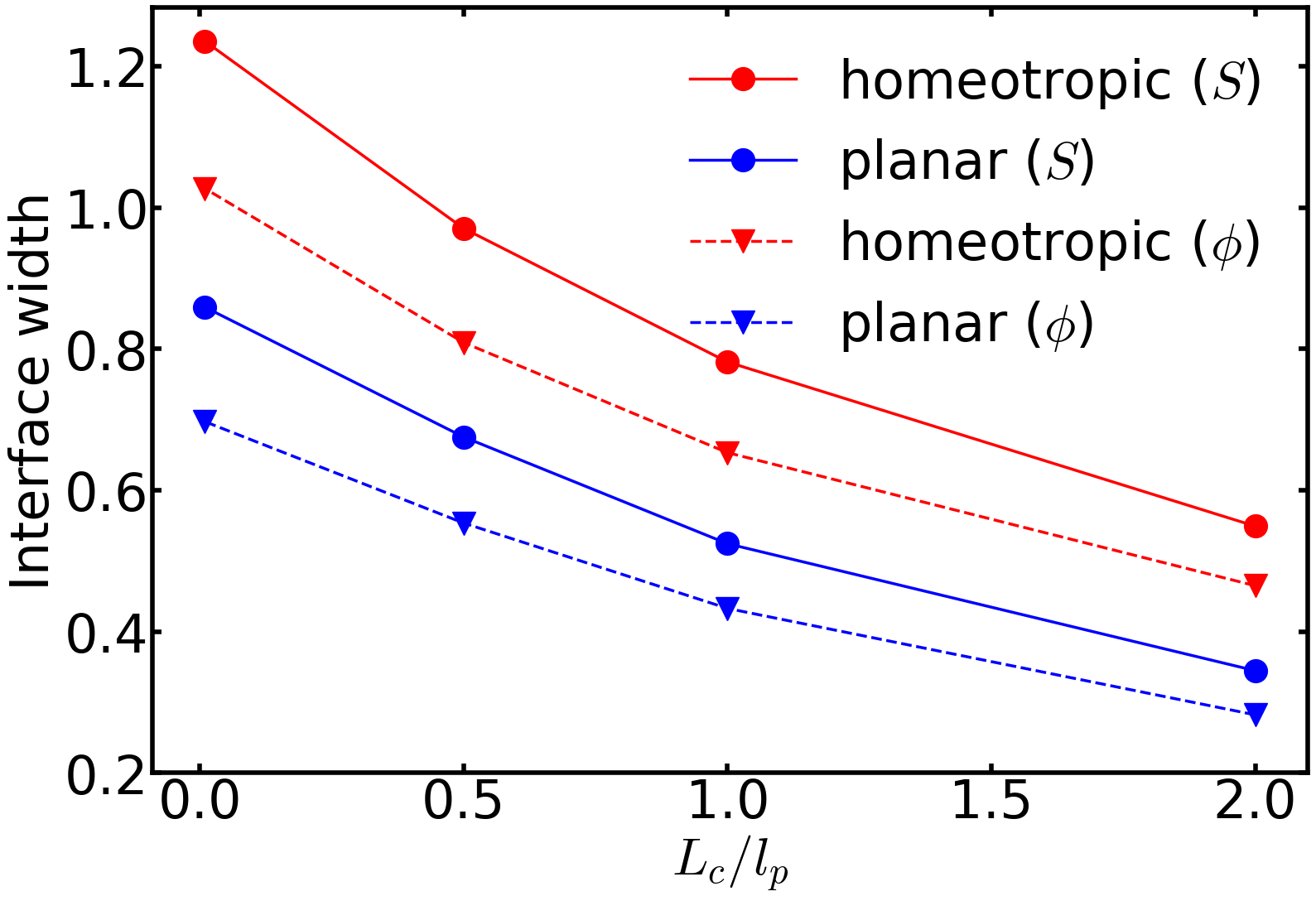}
    \caption{}
    \label{fig:S-u2-nV}
\end{subfigure}
\hfill
\begin{subfigure}{0.4\textwidth}
 \includegraphics[width=\textwidth]{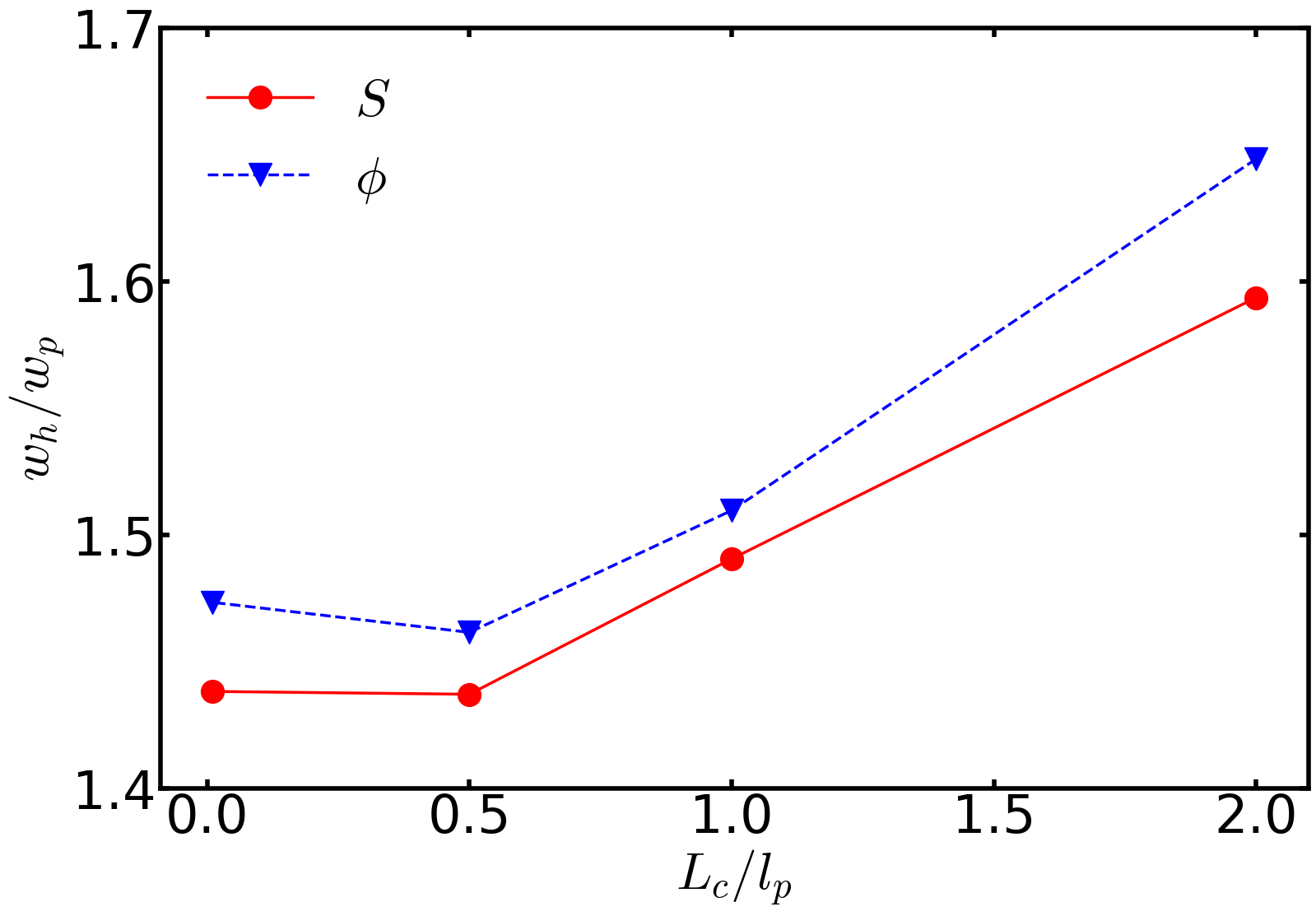}
    \caption{}
    \label{fig:S-u2-lp}
\end{subfigure}
 \hfill
\caption{Interface width dependence on flexibility. For different values of ratio $L/l_p$, the value of $u_2$ is chosen to establish a stable interface near $x=5$ ($u_2=16.1$ for $L_c/l_p=2$, $u_2=9.88$ for $L_c/l_p=1$, $u_2=7.02$ for $L_c/l_p=0.5$ and $u_2=4.58$ for $L_c/l_p=0.01$). (a) shows the widths extracted from both the $\phi$ and $S$ profiles as a function of $L_c/l_p$ for homeotropic and planar alignments. (b) shows the ratio of widths between homeotropic alignment and planar alignment as a function $L_c/l_p$}  
\label{fig:lp}
\end{figure*}

First, we study the interface with homeotropic alignment, where the director $\hat{\mathbf{n}}=\hat{\mathbf{x}}$. Density profiles are plotted in Fig.\ref{fig:u0-u2}(a) for $u_2=9.88$, $10$ and $10.2$. The local density of the isotropic region $\phi_I$ is lower than that of the nematic region, $\phi_N$, leading to a density gap between the nematic and isotropic regions. This result is consistent with previous studies \cite{chen1992numerical, cui1995isotropic, jiang2010isotropic}. In the canonical ensemble, with fixed average chain number density $n/V$, there exists a range of $u_2$ values for which the isotropic and nematic phases coexist. The relative volume fraction of the nematic and isotropic regions depends on the value of $u_{2}$. For higher values of $u_{2}$, the nematic region becomes larger but exhibits a lower density, as shown in Fig. \ref{fig:u0-u2}(a). The shift in the interface position is also reflected in the uniaxial order profile, as depicted in Fig. \ref{fig:u0-u2}(b). In order to extract the width ($w_i$) and location of the interface from our SCFT solutions, we approximate the interfacial uniaxial order profiles by \cite{popa1997statics}
\begin{equation}
S(x) = \frac{S_N}{2}[\tanh(\frac{x-x_0}{w_i})+1]
\end{equation}
where $x_0$ denotes the center of the interface, and $S_N$ represents the uniaxial order in the nematic region. The interfacial widths from the density profiles can be extracted in a similar way. We find that $x_0$ from the density profiles is larger than that from the $S$ profiles, indicating that the density jump is displaced towards the nematic side \cite{jiang2010isotropic}. Additionally, the interfacial width extracted from the density profiles are smaller than that from the $S$ profiles.

We next address how the fluid compressibility, determined by the value of the excluded volume coefficient $u_0$, affects the interface. By setting $u_2=9.88$, the interfaces for different values of $u_0$ are shown in Fig.\ref{fig:u0-u2}(c)(d). As $u_0$ increases, the system becomes more incompressible, leading to a reduction in the gaps of both the uniaxial order and the density profiles. The reduction of $S_N$ widens the interface because segregation between nematic and isotropic phases is smaller \cite{drovetsky1999nematic}. In the limit of infinite $u_0$ (a completely incompressible fluid), the density gaps vanishes, resulting in a uniform density throughout the space, and the interface location is arbitrary. 

Interfacial profiles depend on the director angle in the nematic region relative to the interface normal. The results for planar alignment $\hat{\mathbf{n}}=\hat{\mathbf{z}}$ are shown in Fig. \ref{fig:planar}. Unlike the homeotropic alignment case just described, planar alignment is accompanied by nonzero biaxiality near the interface. Near the interface, but on the isotropic side, the component of the tensor order parameter $Q_{xx}$ is negative, and $Q_{yy} \approx Q_{zz} >0$. This is in agreement with predictions from the Landau-de Gennes free energy \cite{popa1997statics}.

The interfacial width depends on the values of the elastic constants, which, in this model, depends on the flexibility of the chain. Here, we consider systems that are close to the incompressible limit by setting $u_0=50$. Interfacial widths as a function of chain flexibility $L_c/l_p$ are shown in Fig.\ref{fig:lp}(a). Both widths extracted from $\phi$ and $S$ profiles decrease as the chains become more flexible. In the rigid chain limit ($l_p \gg L_c$), the interface width is on order of $L_c$. However, when the chains become flexible, the width scales with $l_p$ instead.

\begin{figure*}[t]
\captionsetup[subfigure]{justification=Centering}
\hfill
\begin{subfigure}{0.23\textwidth}
 \includegraphics[width=\textwidth]{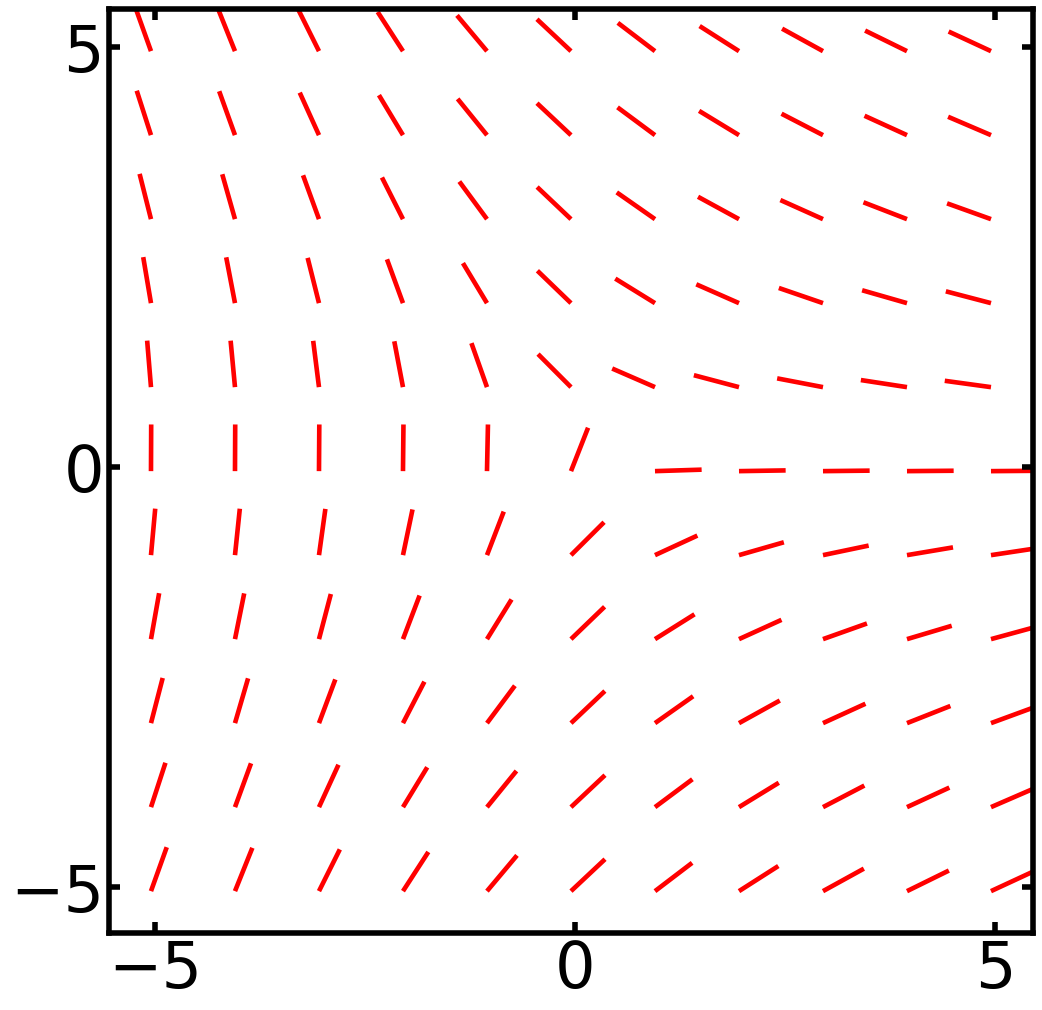}
    \caption{}
    \label{fig:director-minus}
\end{subfigure}
\hfill
\begin{subfigure}{0.23\textwidth}
 \includegraphics[width=\textwidth]{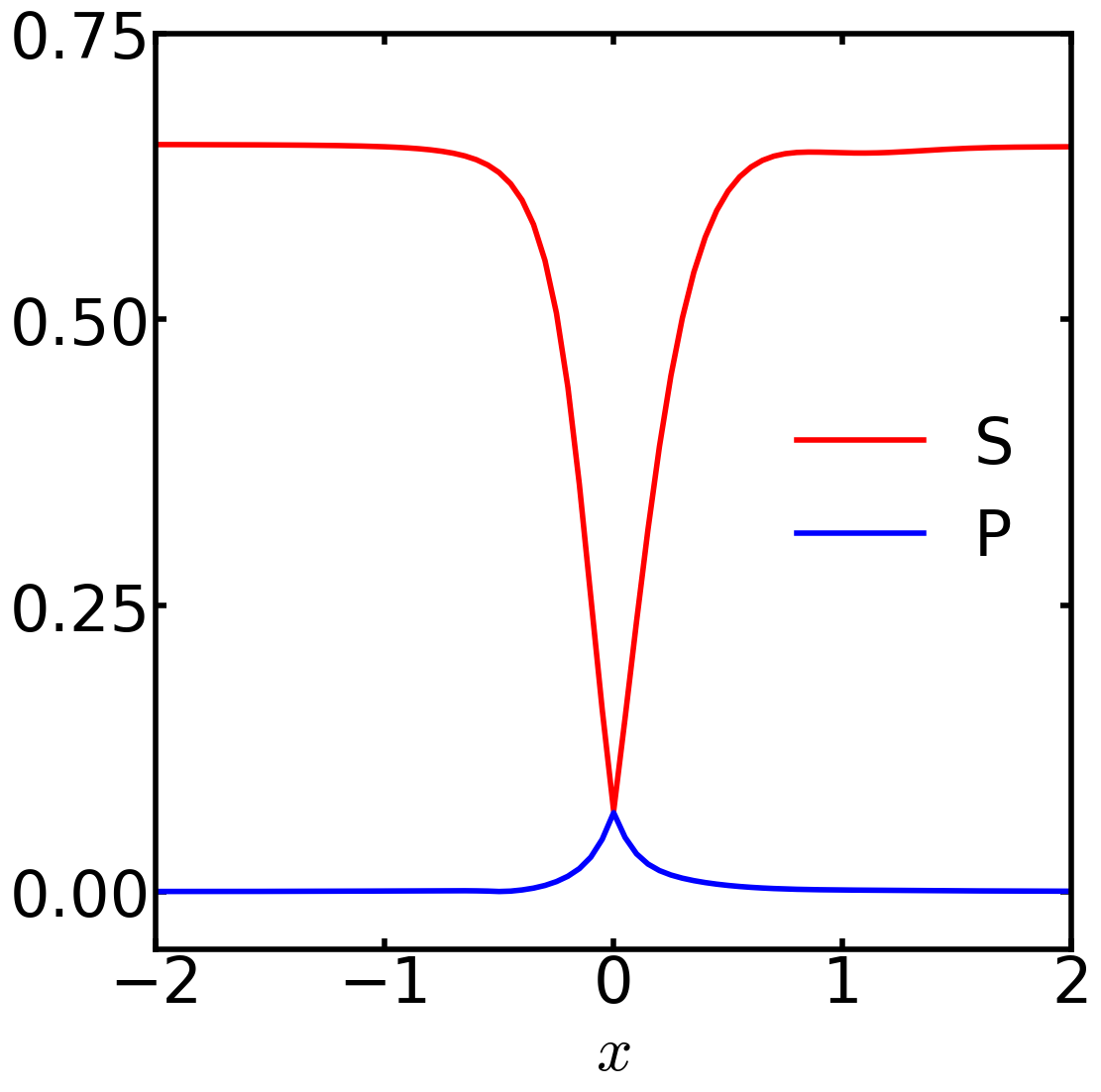}
    \caption{}
    \label{fig:SP-minus}
\end{subfigure}
\hfill
\begin{subfigure}{0.3\textwidth}
 \includegraphics[width=\textwidth]{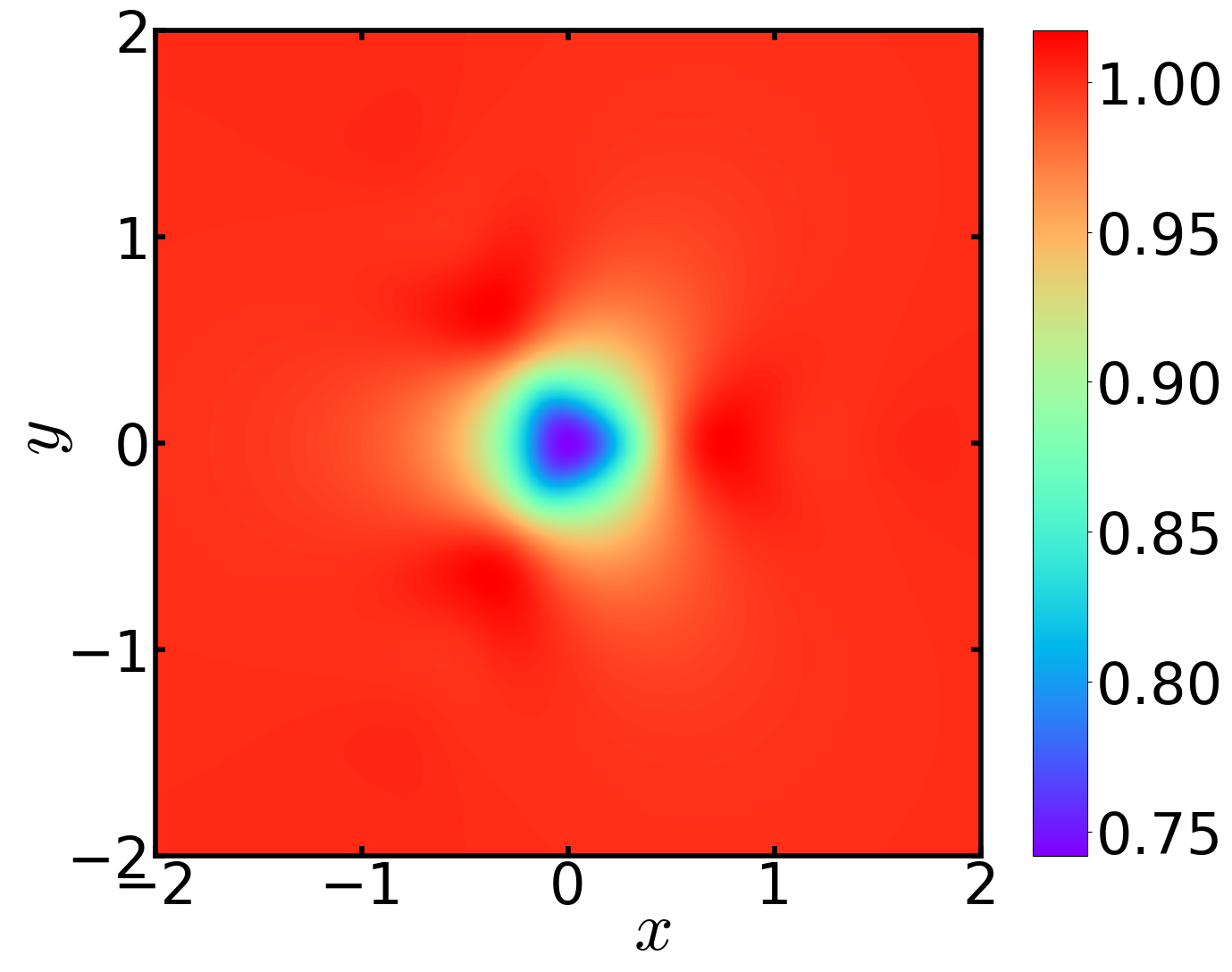}
    \caption{}
    \label{fig:density-minus}
\end{subfigure}
  \hfill

  \hfill
\begin{subfigure}{0.23\textwidth}
 \includegraphics[width=\textwidth]{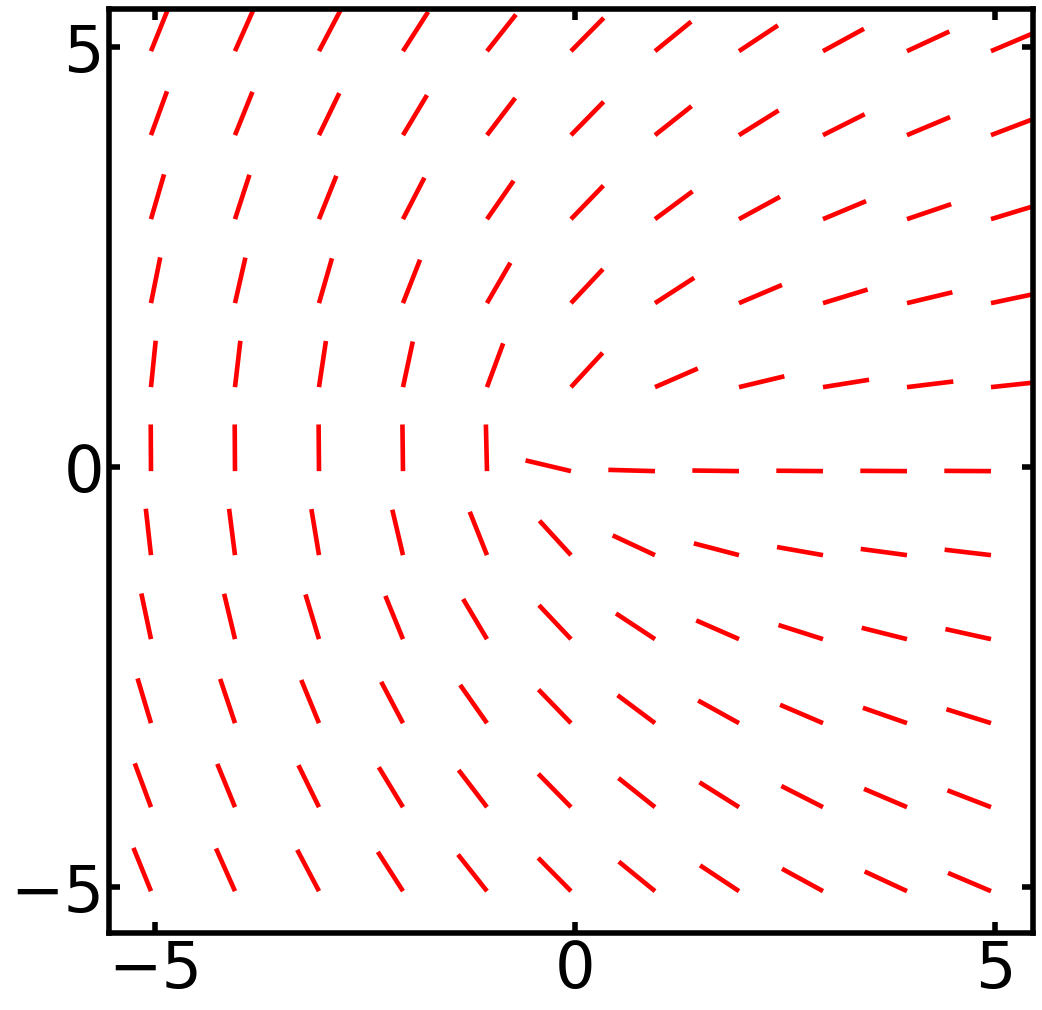}
    \caption{}
    \label{fig:director-plus}
\end{subfigure}
\hfill
\begin{subfigure}{0.23\textwidth}
 \includegraphics[width=\textwidth]{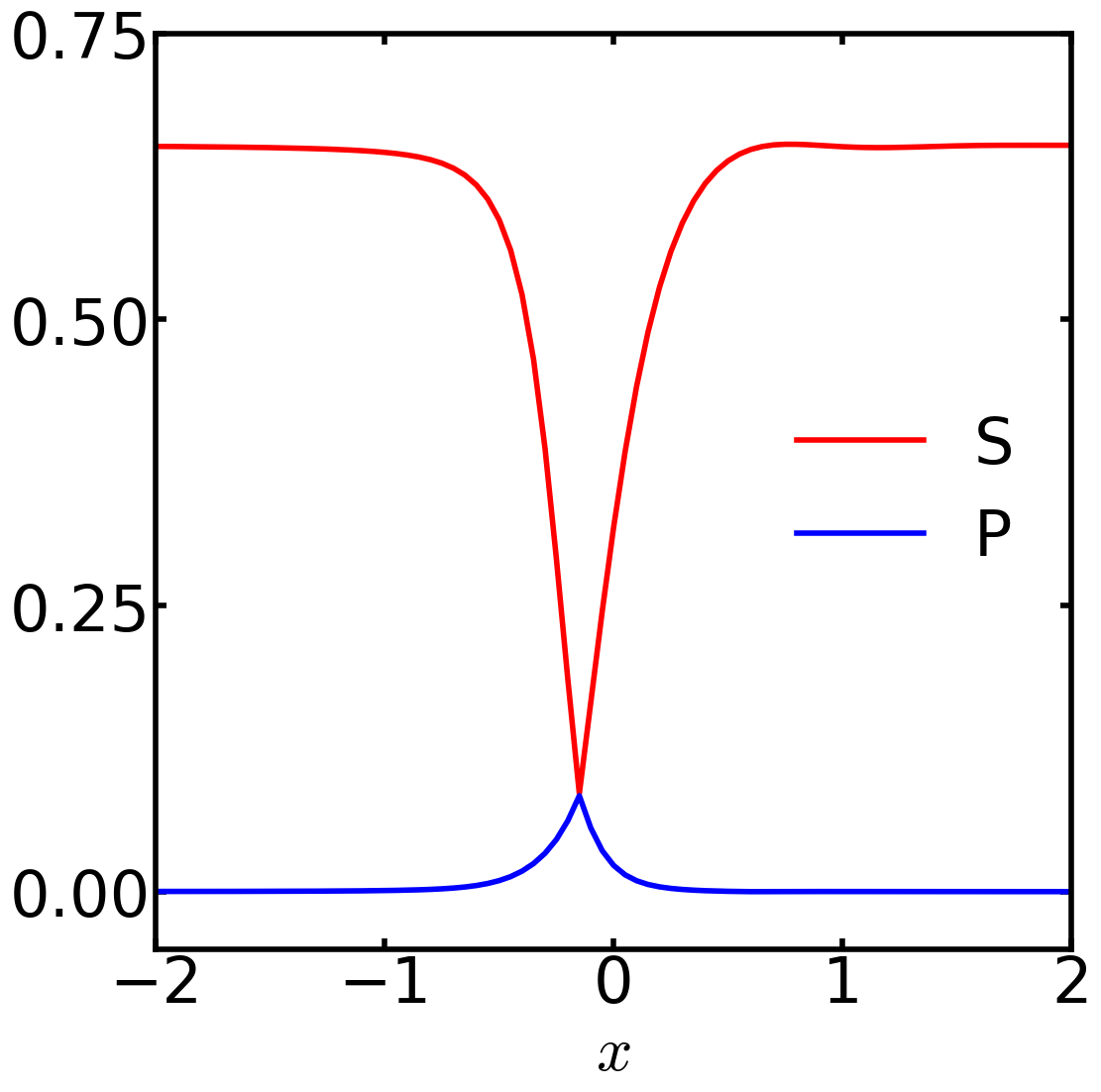}
    \caption{}
    \label{fig:SP-plus}
\end{subfigure}
\hfill
\begin{subfigure}{0.3\textwidth}
 \includegraphics[width=\textwidth]{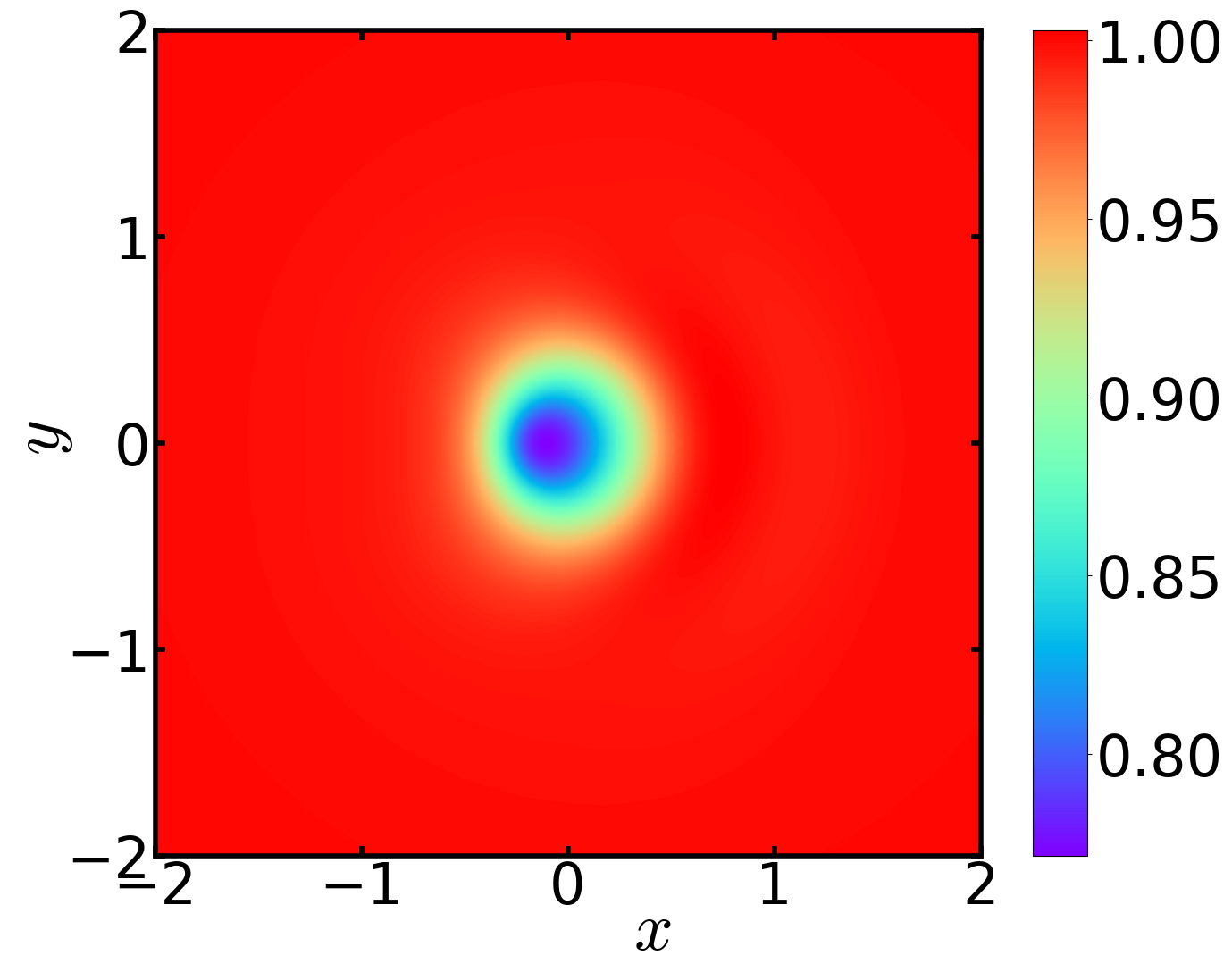}
    \caption{}
    \label{fig:density-plus}
\end{subfigure}
\hfill

\caption{SCFT solutions of $\pm 1/2$ defects. Top: (a), (b), and (c) for -1/2 defect. Bottom: (d), (e), and (f) for 1/2 defect. Left: director field obtained from the largest eigenvalue of $\mathbf{Q}$. Middle: uniaxial (S) and biaxial (P) order parameters as a function of position along the line $y=0$. Right: density field.}  
\label{fig:defect_1}
\end{figure*}

In semiflexible polymer chains, $K_{1}$ is the same order as $K_{3}$, whereas $K_{2}$ can be one order of magnitude smaller than both depending on the flexibility \cite{varytimiadou2024elasticity}. When $K_1 \approx K_3$, the ratio of homeotropic to planar width is predicted to be $\frac{w_h}{w_p}=\sqrt{\frac{6+4\kappa}{6+\kappa}}$, with $\kappa=2(K_1/K_2-1)$ being a measure of the anisotropy \cite{popa1997statics}. In agreement with existing phenomenology, the interfacial width in the case of homeotropic alignment is larger than that of planar alignment \cite{jiang2010isotropic}. The ratio $w_h/w_p$ as a function of $L_c/l_p$ is shown in Fig.\ref{fig:lp}(b); $w_h/w_p$ increases as the chains become very flexible, indicating a higher elastic anisotropy $\kappa$. 

\section{Topological Defects in Two Dimensions}
\label{sec:defects}

We focus here on a different type of non uniform configuration, a topological defect in a two-dimensional square region of lateral dimensions $L_x = L_y = 10$. The region is discretized into $200 \times 200$ uniform square domains. Unless stated otherwise, we set $u_0=15$, $u_2 = 15$, $L_c=l_p = 1$, and $n/V=1$. For the analysis below, we also define polar coordinates $(r, \varphi)$ relative to the center of the defect. The defect center is defined as the location where $S=P$, and it is determined as part of the free energy minimization. Initially, the disclination center is positioned at the geometric center of the square region. The initial configuration is specified as $S(r)=S_N[1-\exp (-5r)]$, $P(r)=0$, and the director angle $\theta = q \varphi$, where $q = \pm 1/2$ is considered in this work. To obtain the SCFT solution, the iteration process terminates once the maximum error in the $w_l^m$ field falls below $\epsilon=0.01$.

\begin{figure*}[t]
\captionsetup[subfigure]{justification=Centering}
\hfill
\begin{subfigure}{0.26\textwidth}
 \includegraphics[width=\textwidth]{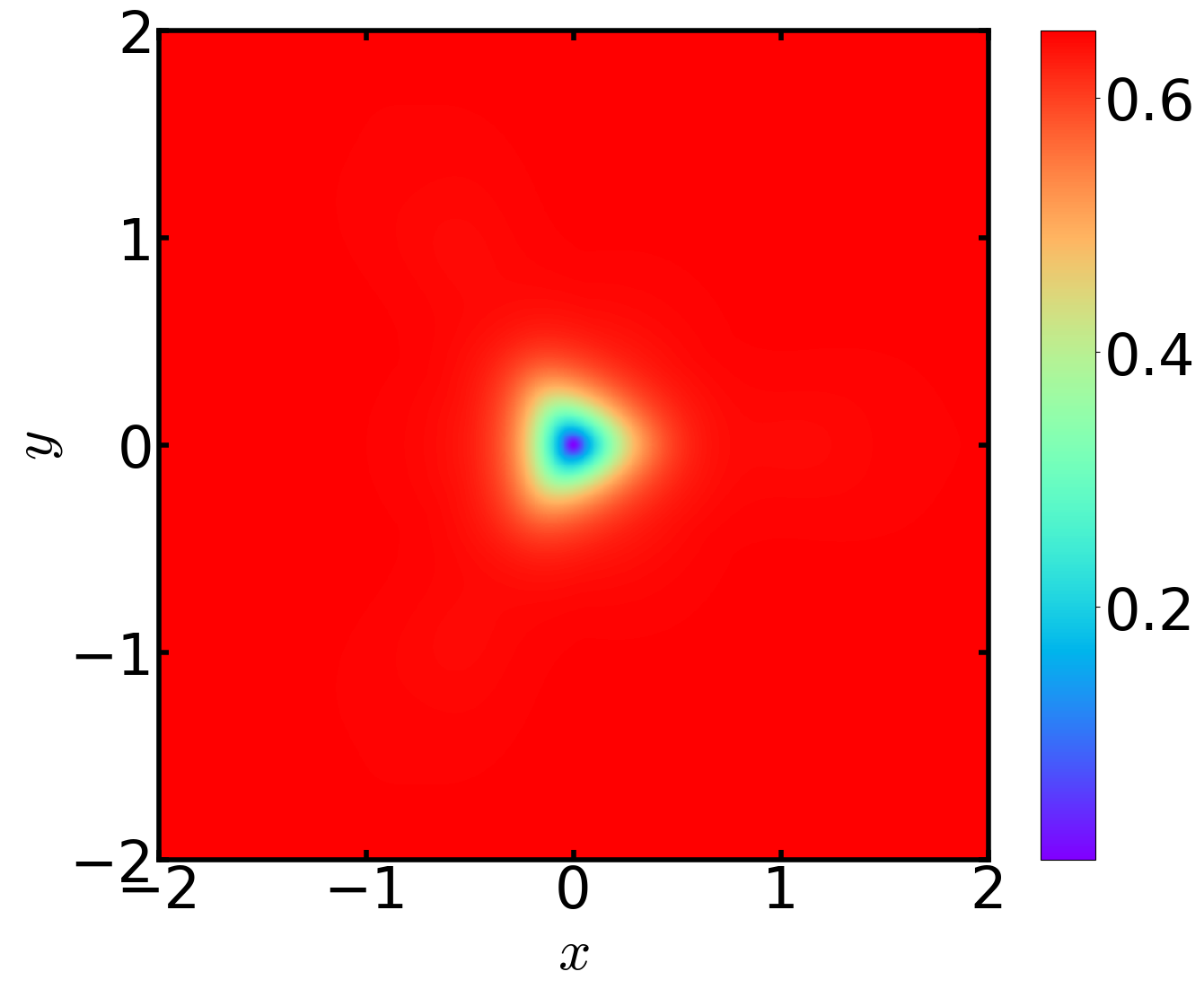}
    \caption{}
    \label{fig:Gamma-minus}
\end{subfigure}
 \hfill
\begin{subfigure}{0.3\textwidth}
 \includegraphics[width=\textwidth]{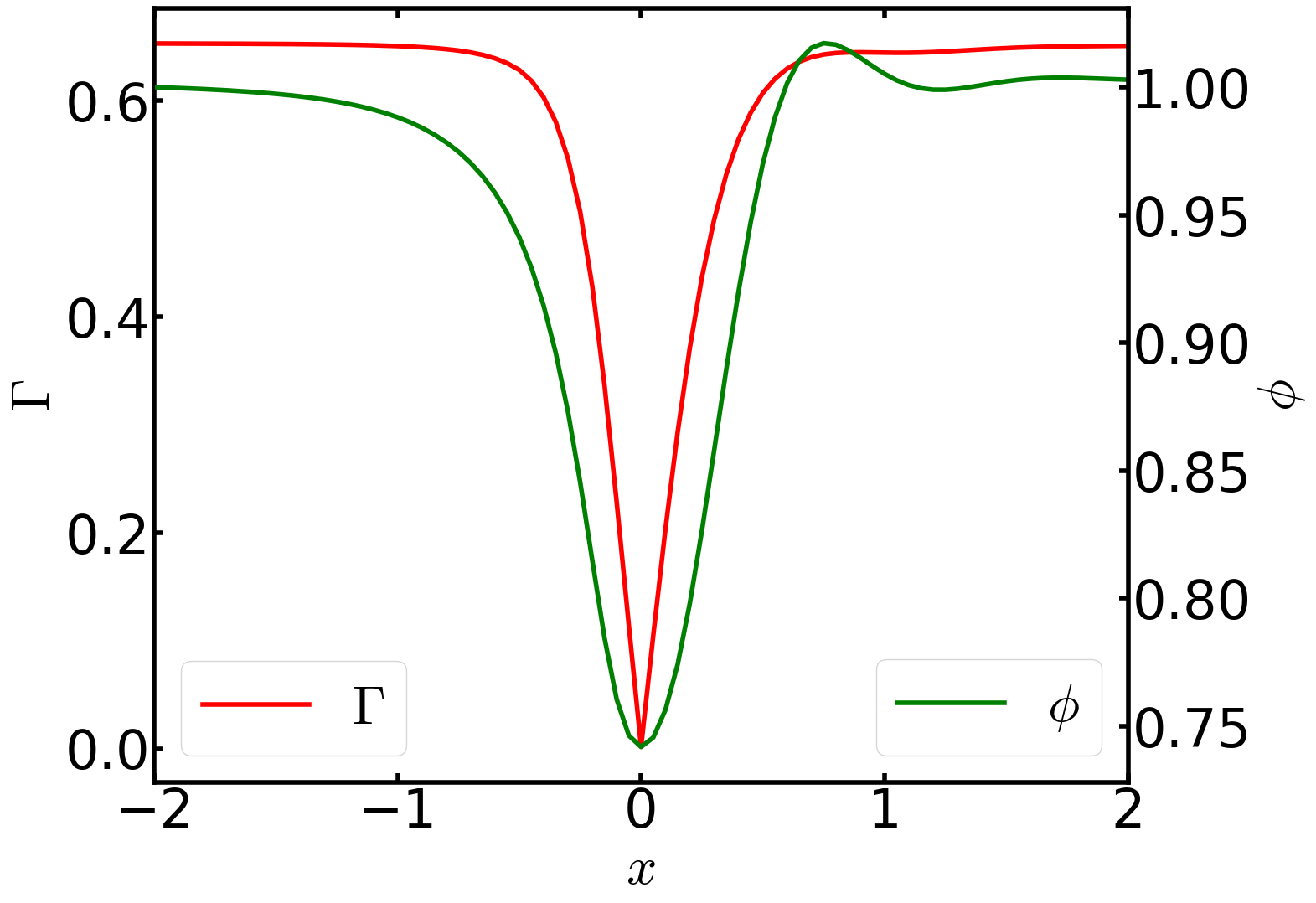}
    \caption{}
    \label{fig:Gamma-density-minus}
\end{subfigure}
 \hfill
\begin{subfigure}{0.28\textwidth}
 \includegraphics[width=\textwidth]{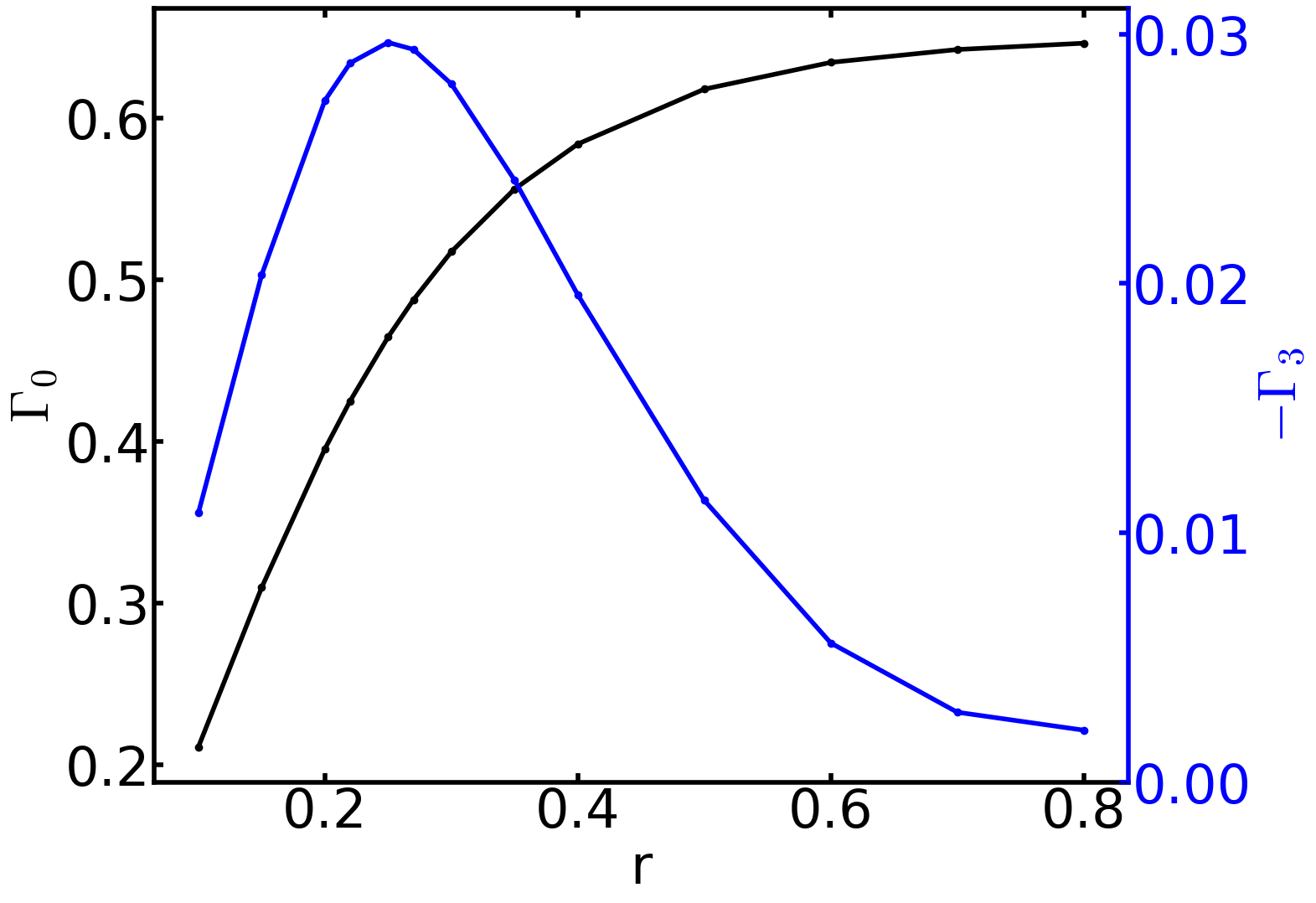}
    \caption{}
    \label{fig:density-minus}
\end{subfigure}
\hfill

\hfill
\begin{subfigure}{0.26\textwidth}
 \includegraphics[width=\textwidth]{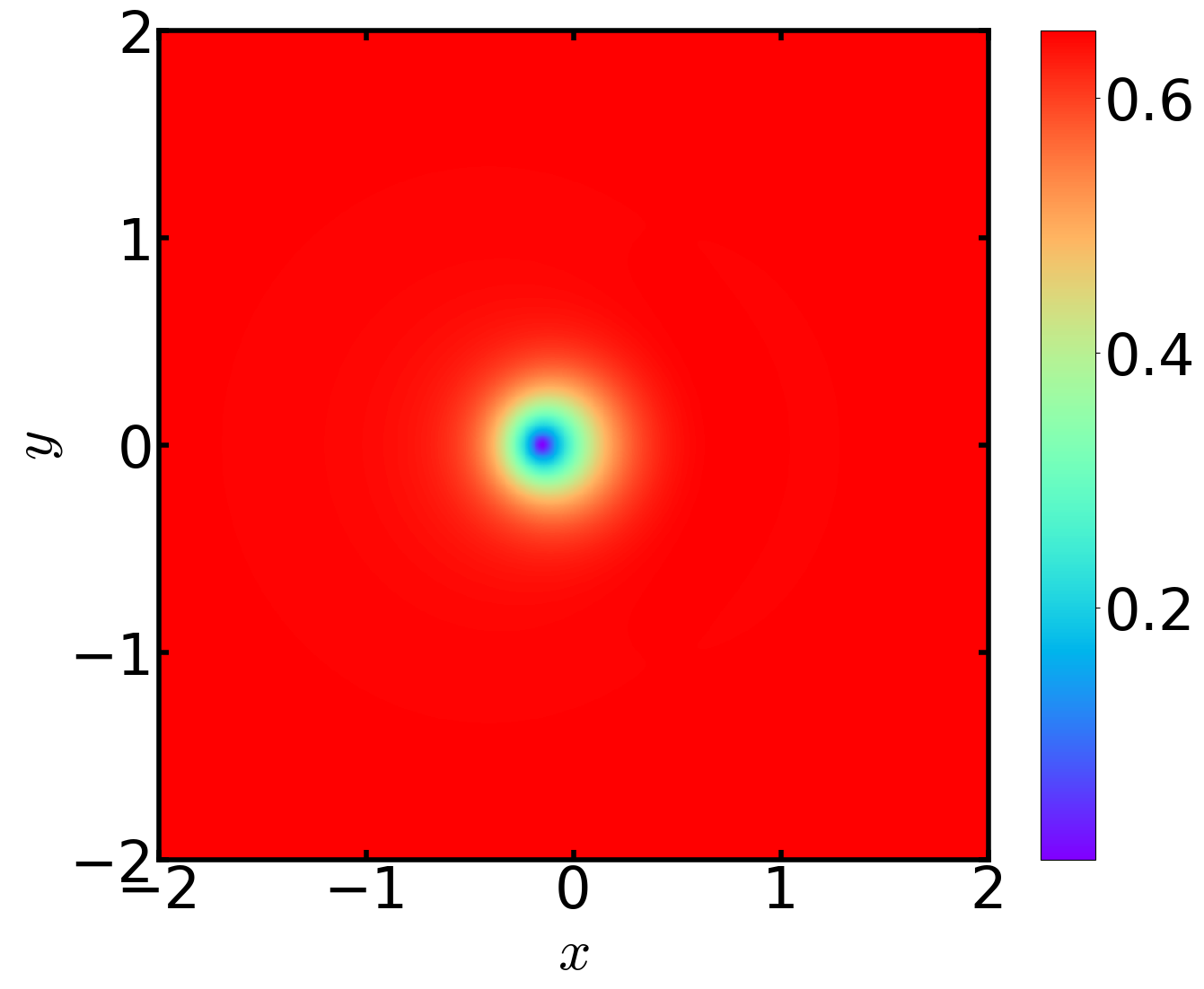}
    \caption{}
    \label{fig:Gamma-plus}
\end{subfigure}
\hfill
\begin{subfigure}{0.3\textwidth}
 \includegraphics[width=\textwidth]{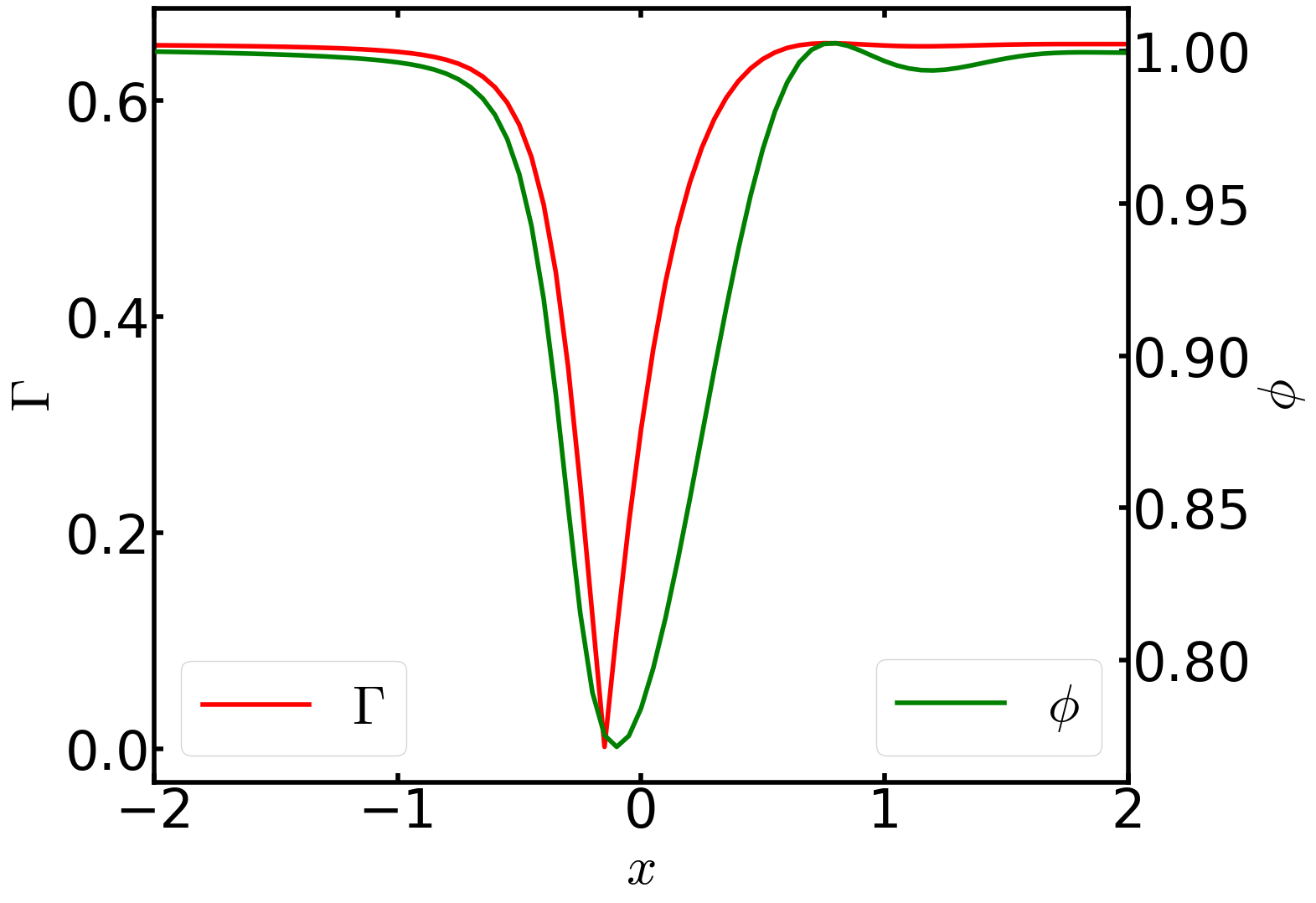}
    \caption{}
    \label{fig:Gamma-density-plus}
\end{subfigure}
 \hfill
\begin{subfigure}{0.28\textwidth}
 \includegraphics[width=\textwidth]{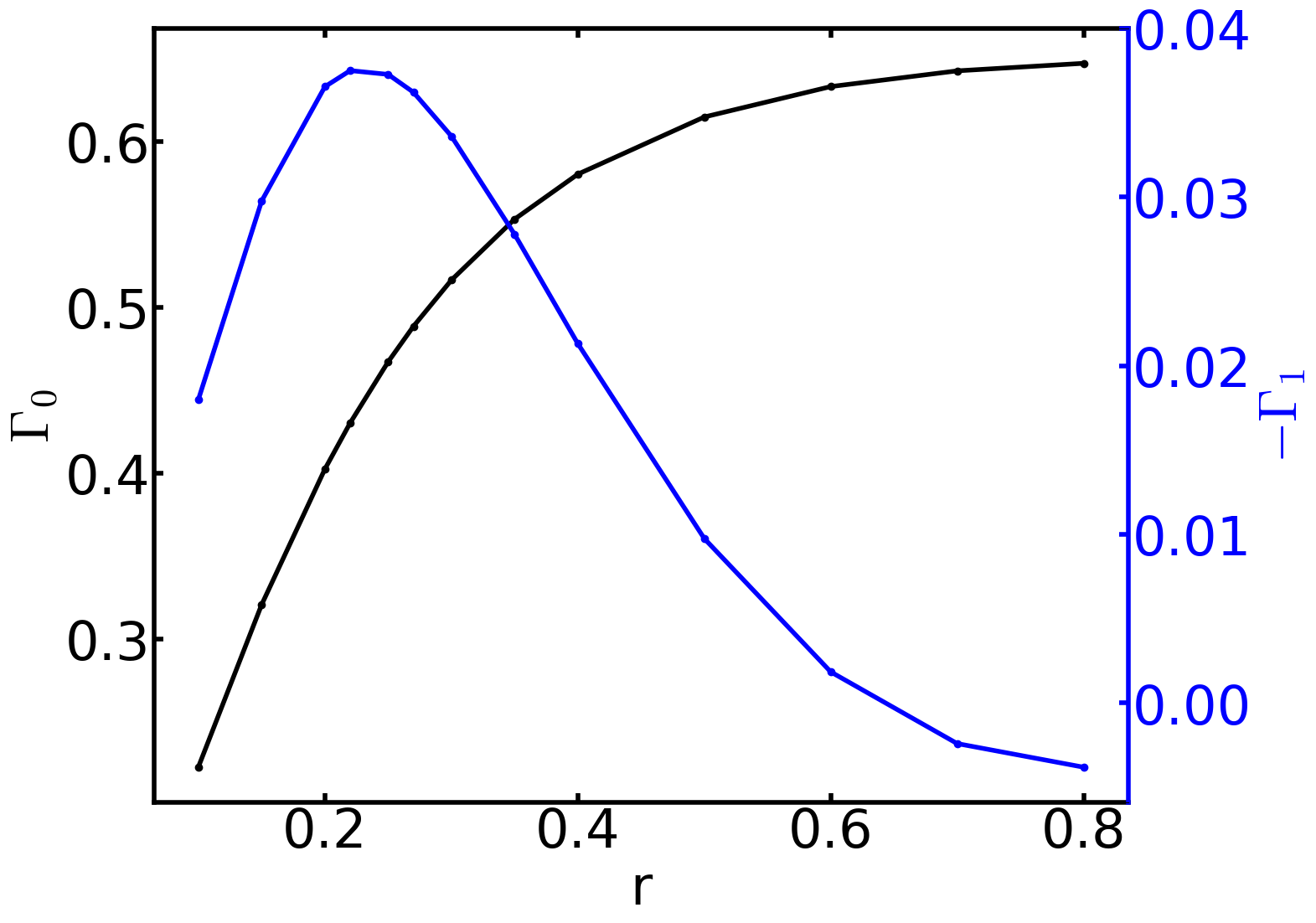}
    \caption{}
    \label{fig:density-plus}
\end{subfigure}
 \hfill
\caption{SCFT solutions for $\pm 1/2$ defects. Top: (a), (b), and (c) correspond to a -1/2 defect. Bottom: (d), (e), and (f) to a 1/2 defect. Left: spatial distribution of $\Gamma$. Middle: $\Gamma$ and the $\phi$ as a function of position along the line $y=0$. Right: Fourier components of $\Gamma$ as a function of the radial distance from the defect center.}  
\label{fig:defect_2}
\end{figure*}

Figure \ref{fig:defect_1} presents the SCFT solutions for $q = \pm 1/2$ defects. The director profiles are shown in Figures \ref{fig:defect_1}(a) and \ref{fig:defect_1}(d) for $q = -1/2$ and $q = 1/2$, respectively. Figures
\ref{fig:defect_1}(b) and \ref{fig:defect_1}(e) illustrate the corresponding uniaxial ($S$) and biaxial ($P$) order parameters along the line $y=0$. For $q=-1/2$, the defect core remains at the geometric center. In contrast, for $q=1/2$, the defect center shifts toward the $-\hat{\mathbf{x}}$ direction as the iterations progress. In both cases, the configurations are  uniaxial ($P = 0$) far from the defect core. However, as the defect core is approached, $P$ increases, leading to a locally biaxial configuration. At the exact center of the defect, where $S=P$, the tensor $\mathbf{Q}$ exhibits two degenerate positive eigenvalues, resulting in a uniaxial configuration once again \cite{re:schimming22,re:schimming23}. Figures \ref{fig:defect_1}(c) (f) show the anisotropic density distributions near the defect cores. In both cases, the core regions exhibit lower densities compared to the surrounding nematic regions. Figures \ref{fig:defect_2}(a) (d) present the distributions of optical retardance, $\Gamma = S-P$, a quantity that can be experimentally measured through polarization microscopy. The $\Gamma$ distributions reveal cores of smaller extent compared to what would be inferred from the density distributions. Figures \ref{fig:defect_2}(b)(e) display the density and optical retardance along the line $y=0$. The density profiles are smooth across the cores, while $\Gamma$ exhibits singular behavior at the cores. It is noteworthy that, in the case of $q=-1/2$, the location of lowest density coincides with the defect core center. However, for $q=1/2$, the location of the lowest density does not align with the defect core center. The $\Gamma$ distributions are further analyzed by their angular Fourier modes $\Gamma(r, \varphi)=\sum \Gamma_n(r) \cos (n\varphi)$, as shown in Fig.\ref{fig:defect_2}(c) (f). The nonzero anisotropic term $\Gamma_1$ in $1/2$ defect and $\Gamma_3$ in $-1/2$ defect are signatures of the anisotropy in splay and bend constants \cite{re:zhou17,schimming2020anisotropic}.

\begin{figure*}[h!]
\captionsetup[subfigure]{justification=Centering}
\hfill
\begin{subfigure}{0.3\textwidth}
 \includegraphics[width=\textwidth]{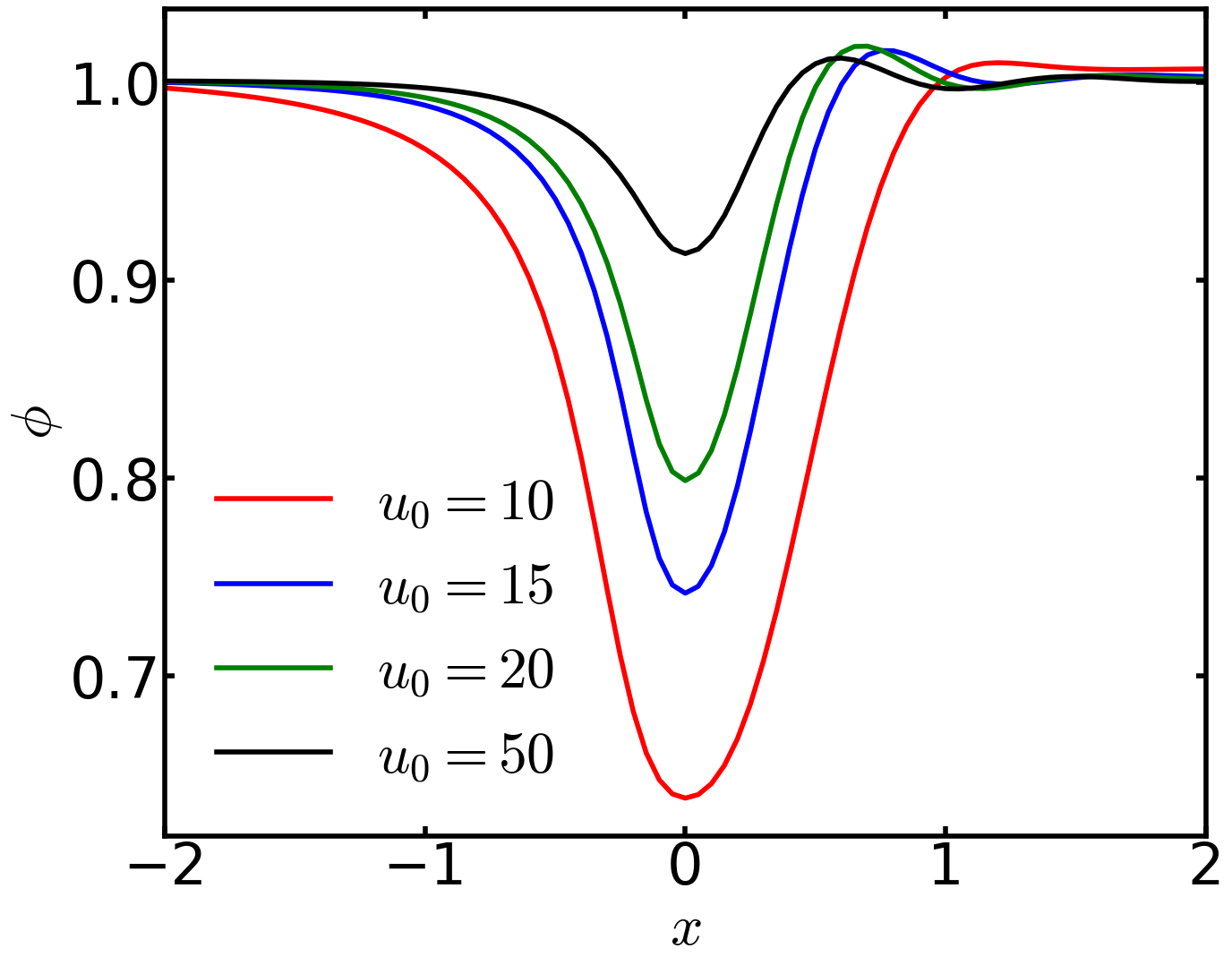}
    \caption{}
    \label{fig:Gamma-minus}
\end{subfigure}
 \hfill
\begin{subfigure}{0.3\textwidth}
 \includegraphics[width=\textwidth]{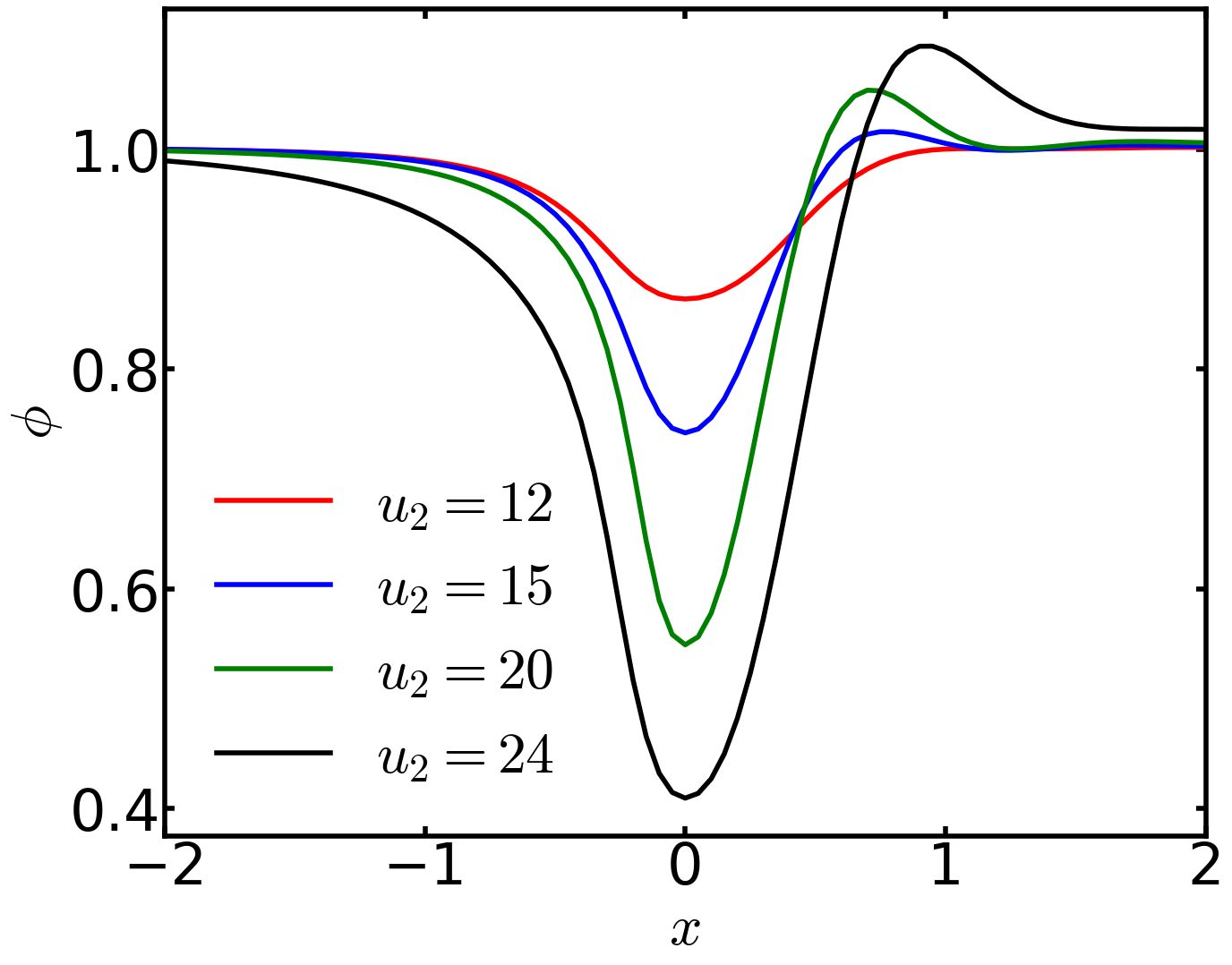}
    \caption{}
    \label{fig:Gamma-density-minus}
\end{subfigure}
 \hfill
\begin{subfigure}{0.3\textwidth}
 \includegraphics[width=\textwidth]{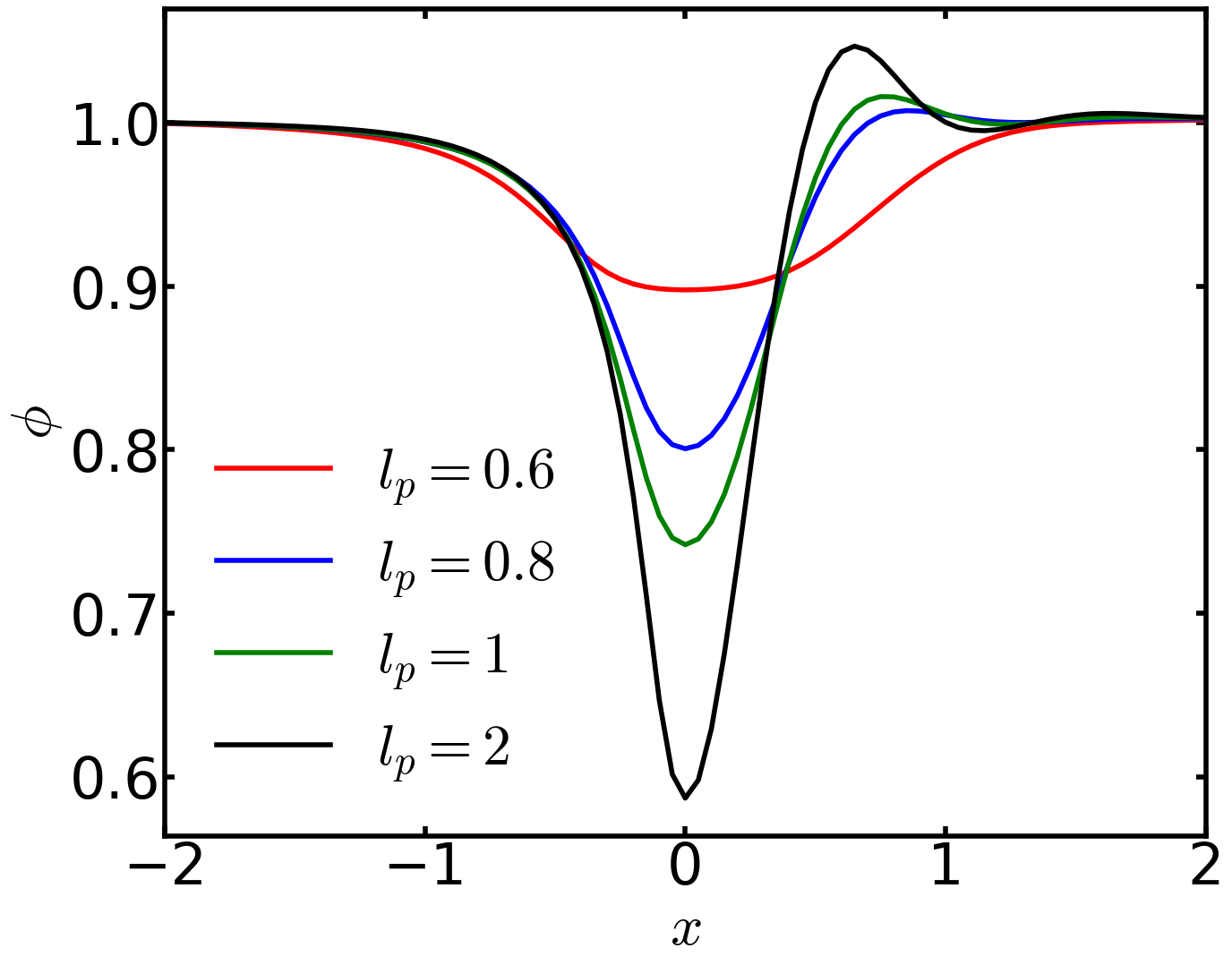}
    \caption{}
    \label{fig:density-minus}
\end{subfigure}
\hfill

\hfill
\begin{subfigure}{0.3\textwidth}
 \includegraphics[width=\textwidth]{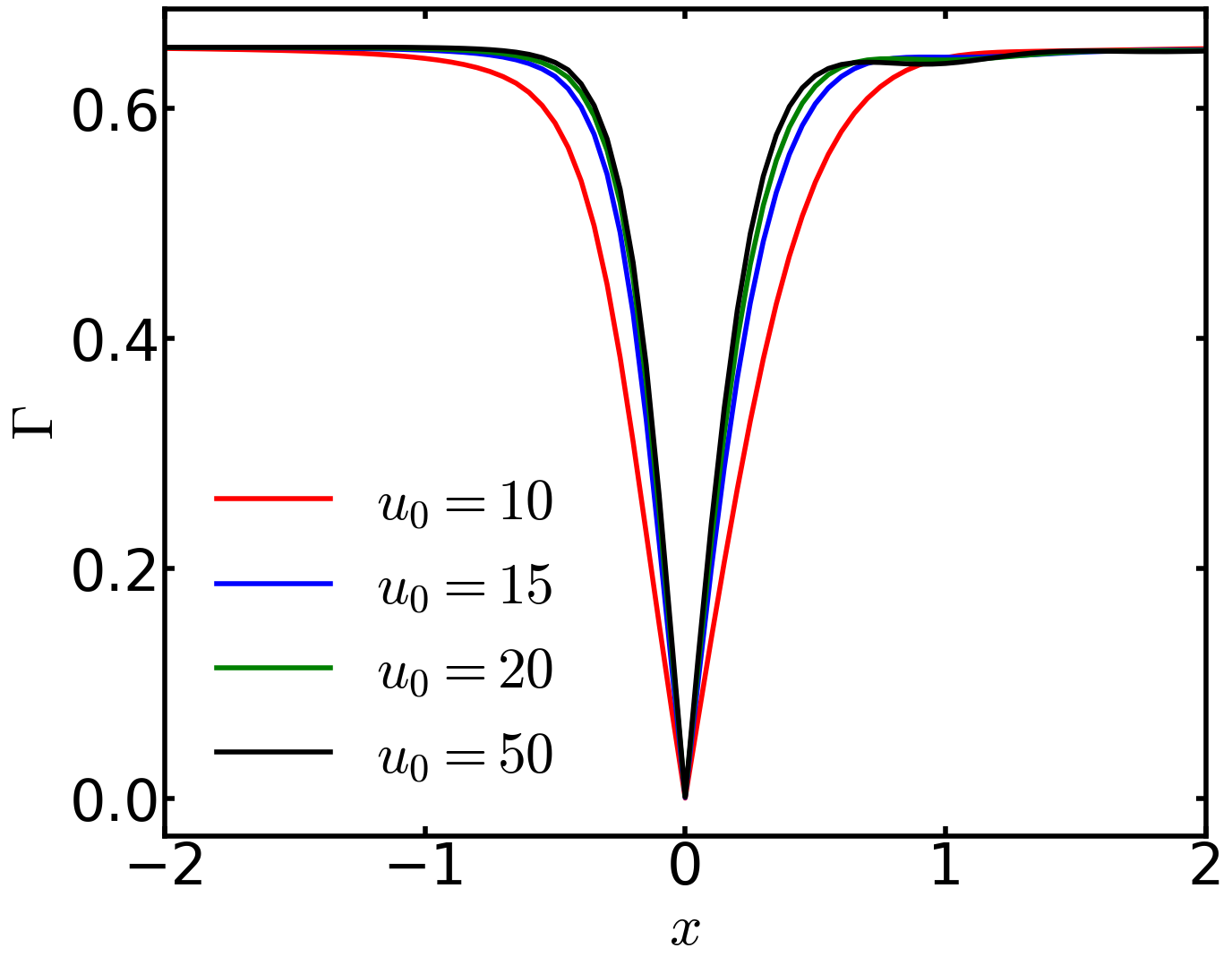}
    \caption{}
    \label{fig:Gamma-plus}
\end{subfigure}
\hfill
\begin{subfigure}{0.3\textwidth}
 \includegraphics[width=\textwidth]{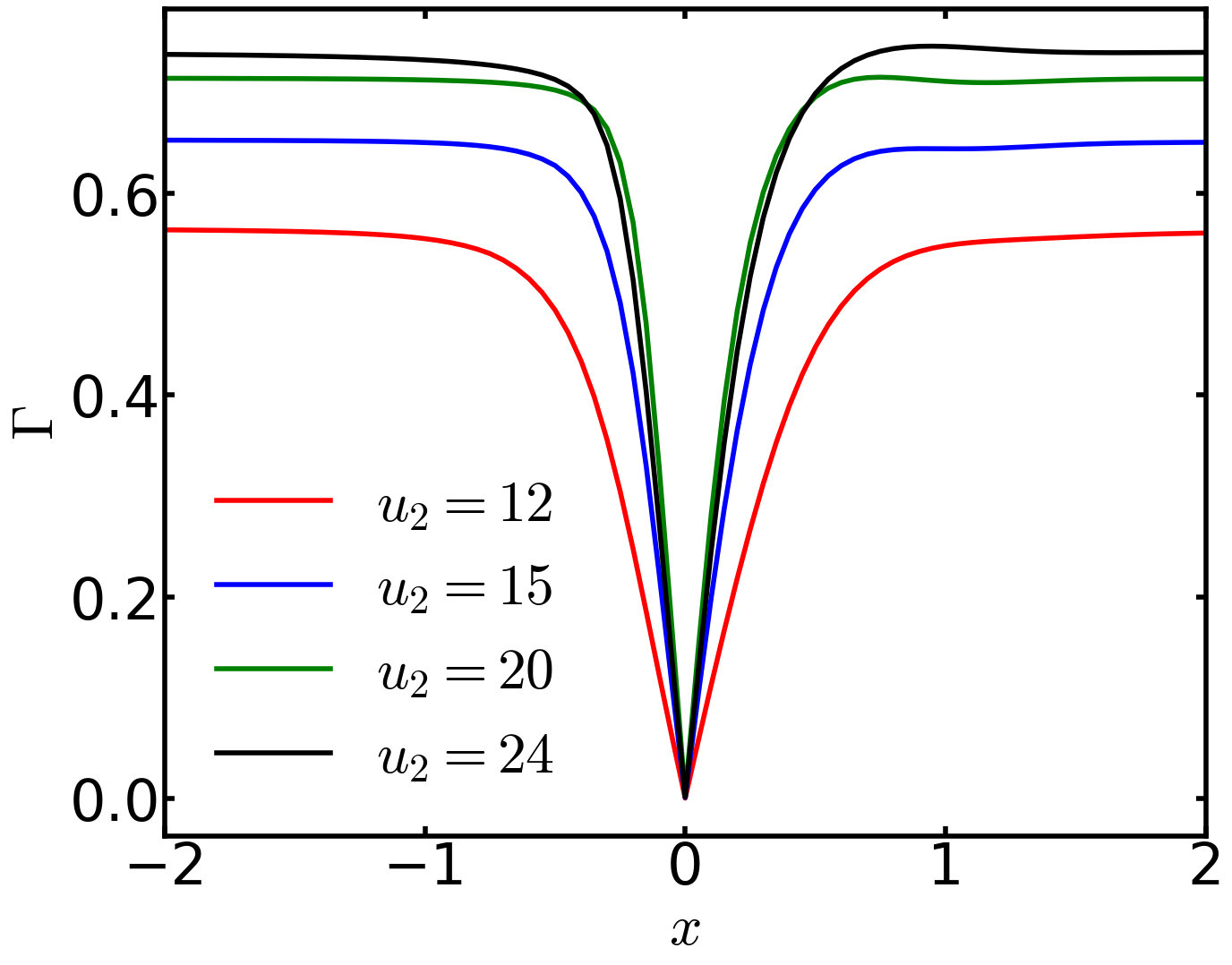}
    \caption{}
    \label{fig:Gamma-density-plus}
\end{subfigure}
 \hfill
\begin{subfigure}{0.3\textwidth}
 \includegraphics[width=\textwidth]{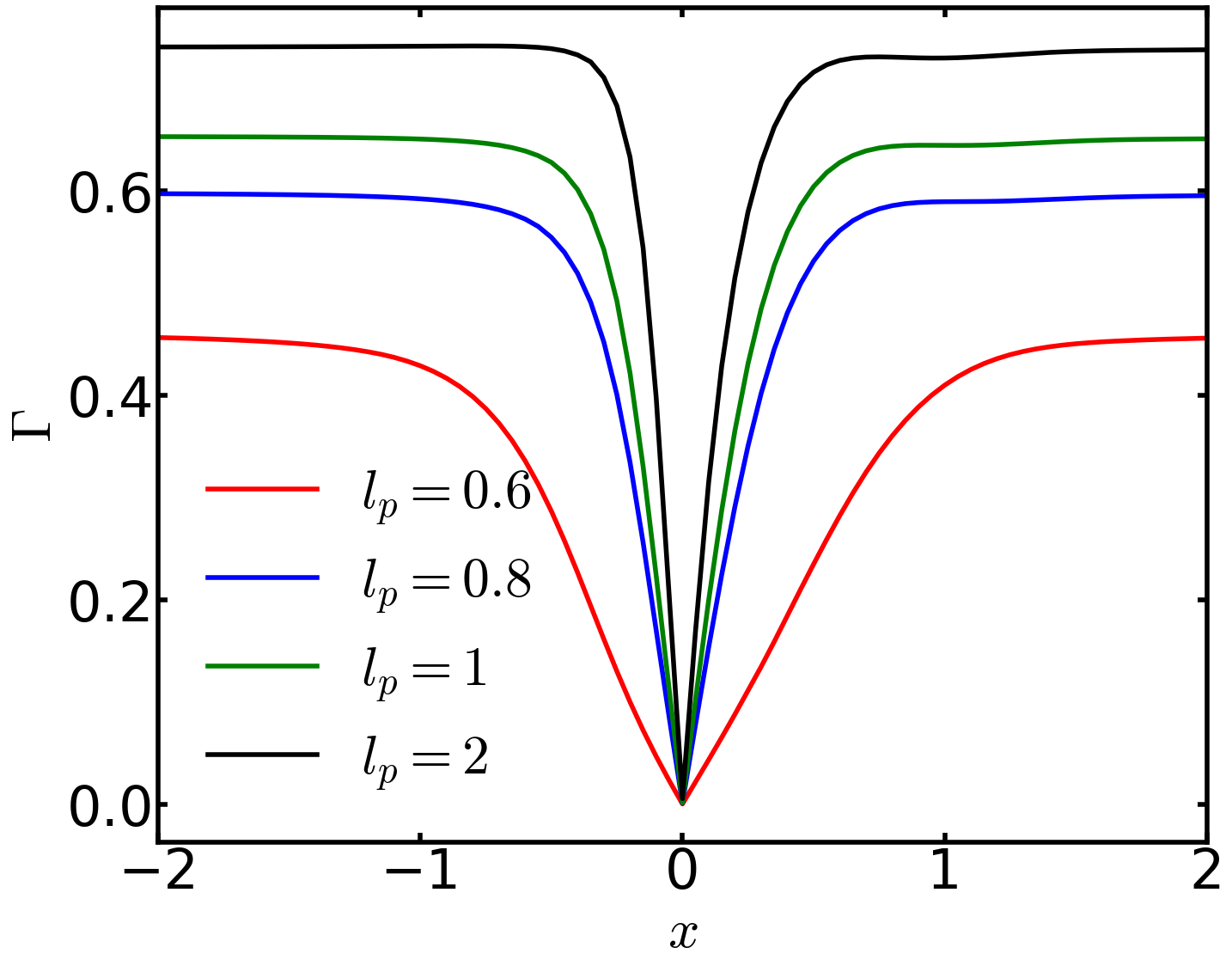}
    \caption{}
    \label{fig:density-plus}
\end{subfigure}
 \hfill
\caption{Density ($\phi$) and optical retardance ($\Gamma$) along the line $y=0$ for $-1/2$ defect at varying values of (a)(d)$u_0$, (b)(e)$u_2$ and (c)(f)$l_p$}  
\label{fig:cores}
\end{figure*}

\begin{figure*}[t]
\captionsetup[subfigure]{justification=Centering}
\hfill
\begin{subfigure}{0.3\textwidth}
 \includegraphics[width=\textwidth]{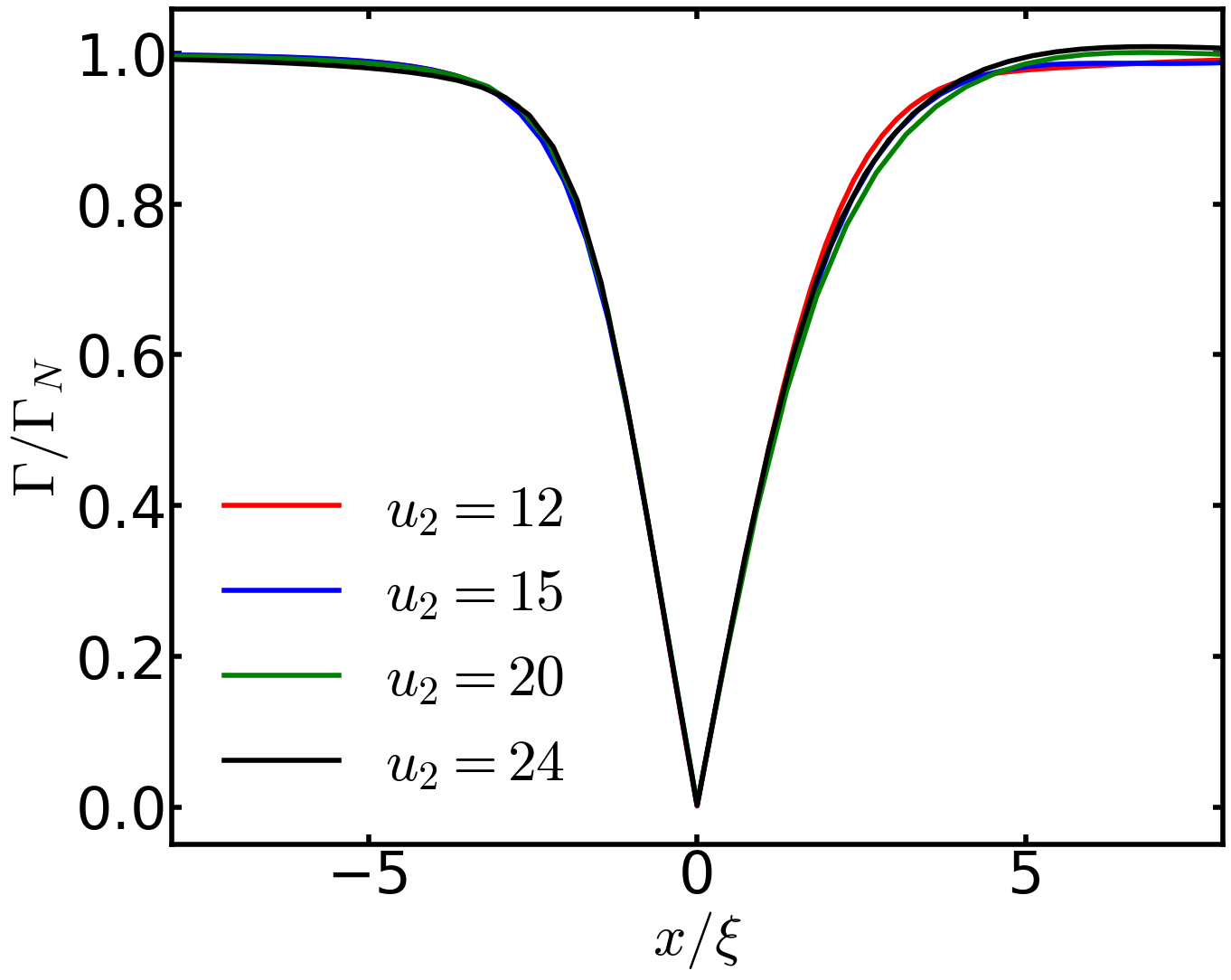}
    \caption{}
    \label{fig:Gamma-u2-normalized}
\end{subfigure}
\hfill
\begin{subfigure}{0.3\textwidth}
 \includegraphics[width=\textwidth]{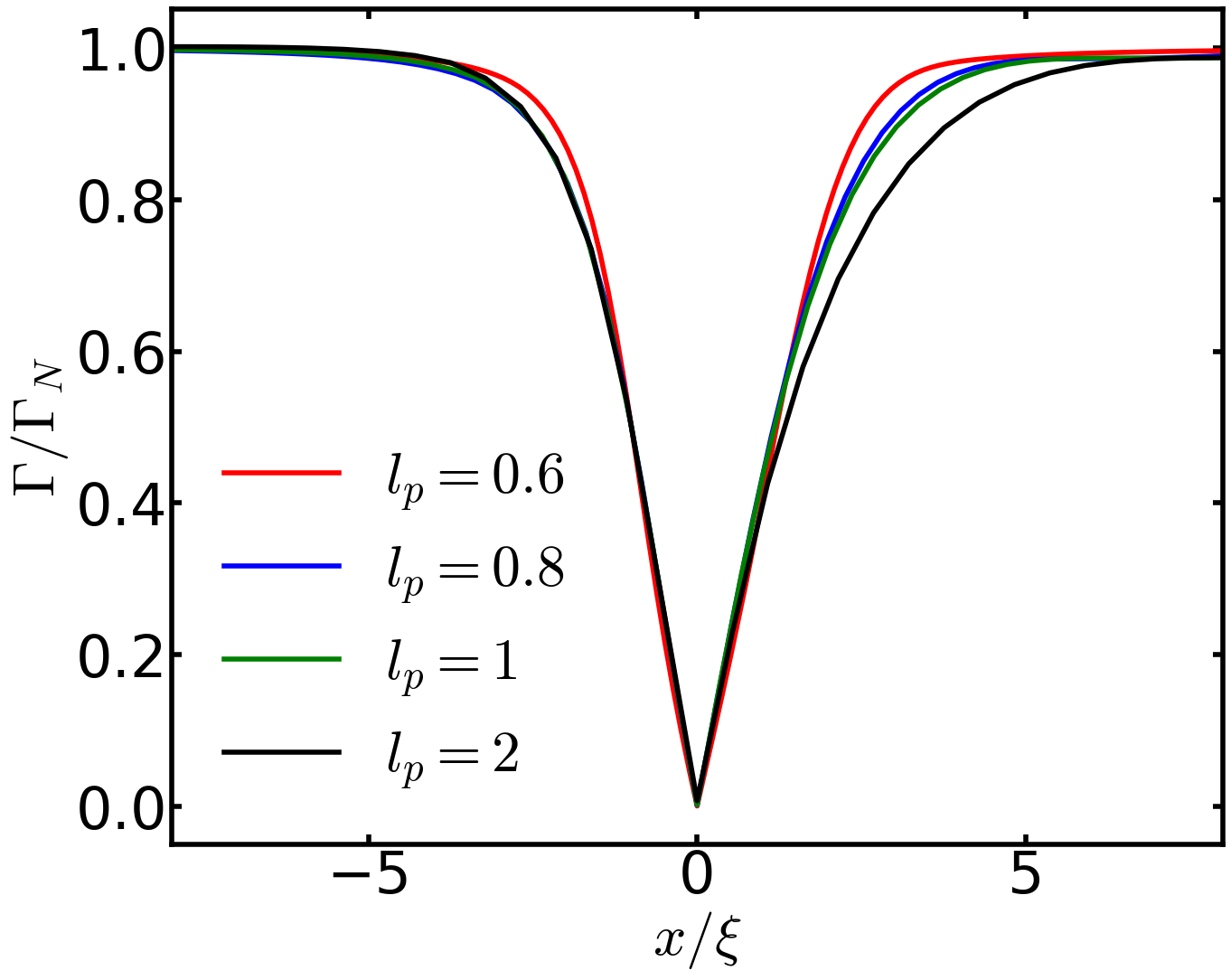}
    \caption{}
    \label{fig:Gamma-lp-normalized}
\end{subfigure}
 \hfill
 \begin{subfigure}{0.3\textwidth}
 \includegraphics[width=\textwidth]{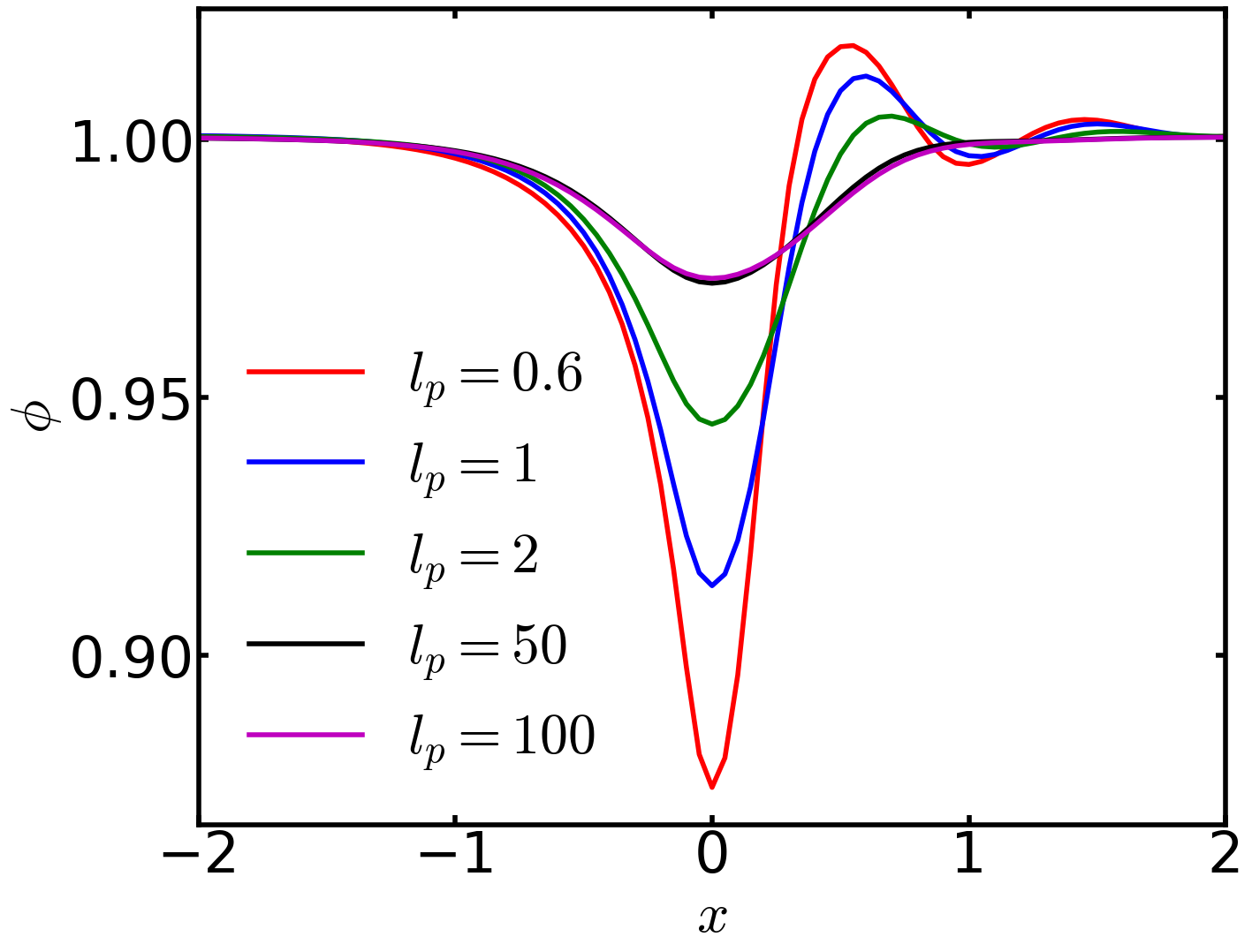}
    \caption{}
    \label{fig:phi-lp-fixedS}
\end{subfigure}
 \hfill

 \hfill
\begin{subfigure}{0.3\textwidth}
 \includegraphics[width=\textwidth]{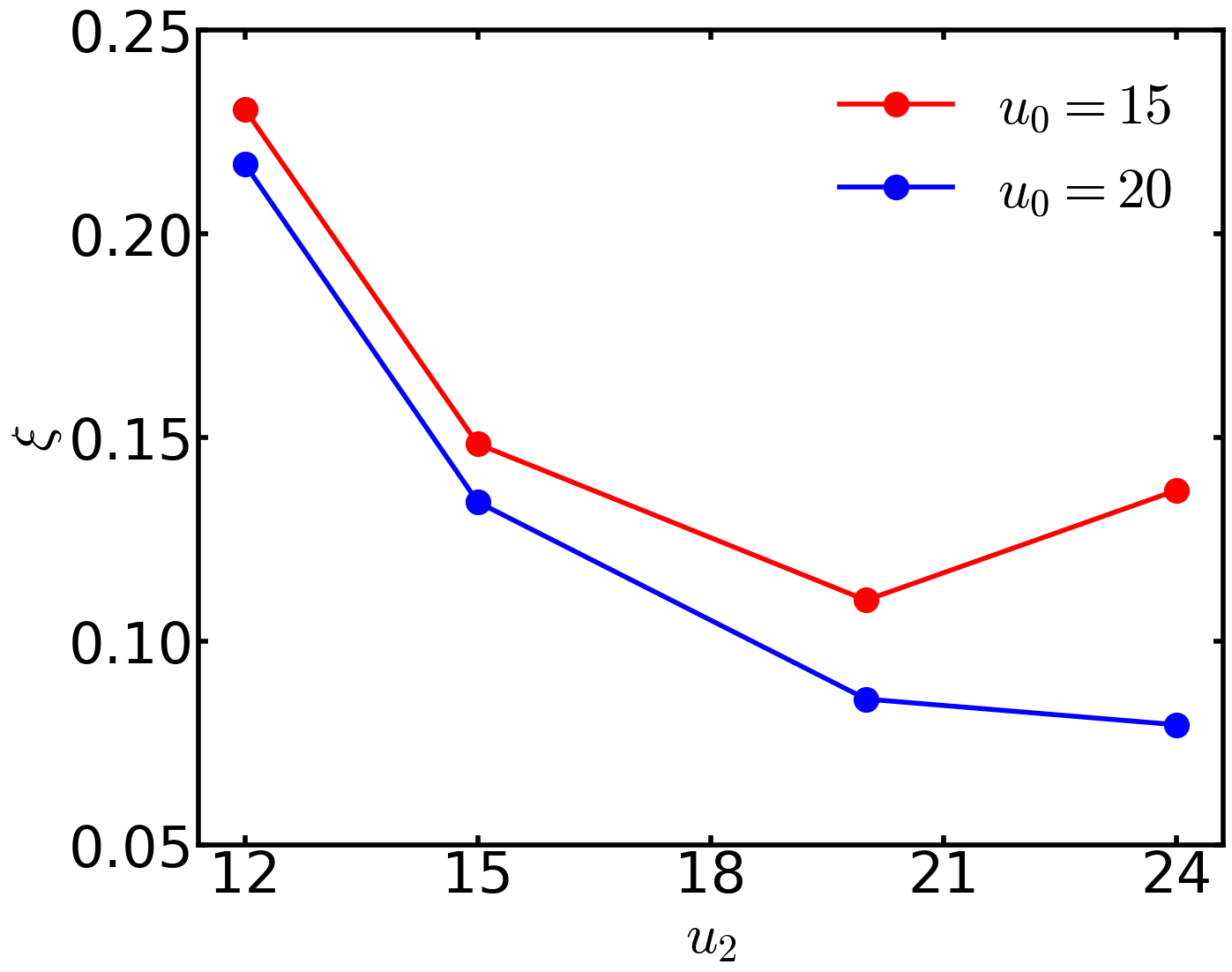}
    \caption{}
    \label{fig:xi-u2}
\end{subfigure}
\hfill
\begin{subfigure}{0.3\textwidth}
 \includegraphics[width=\textwidth]{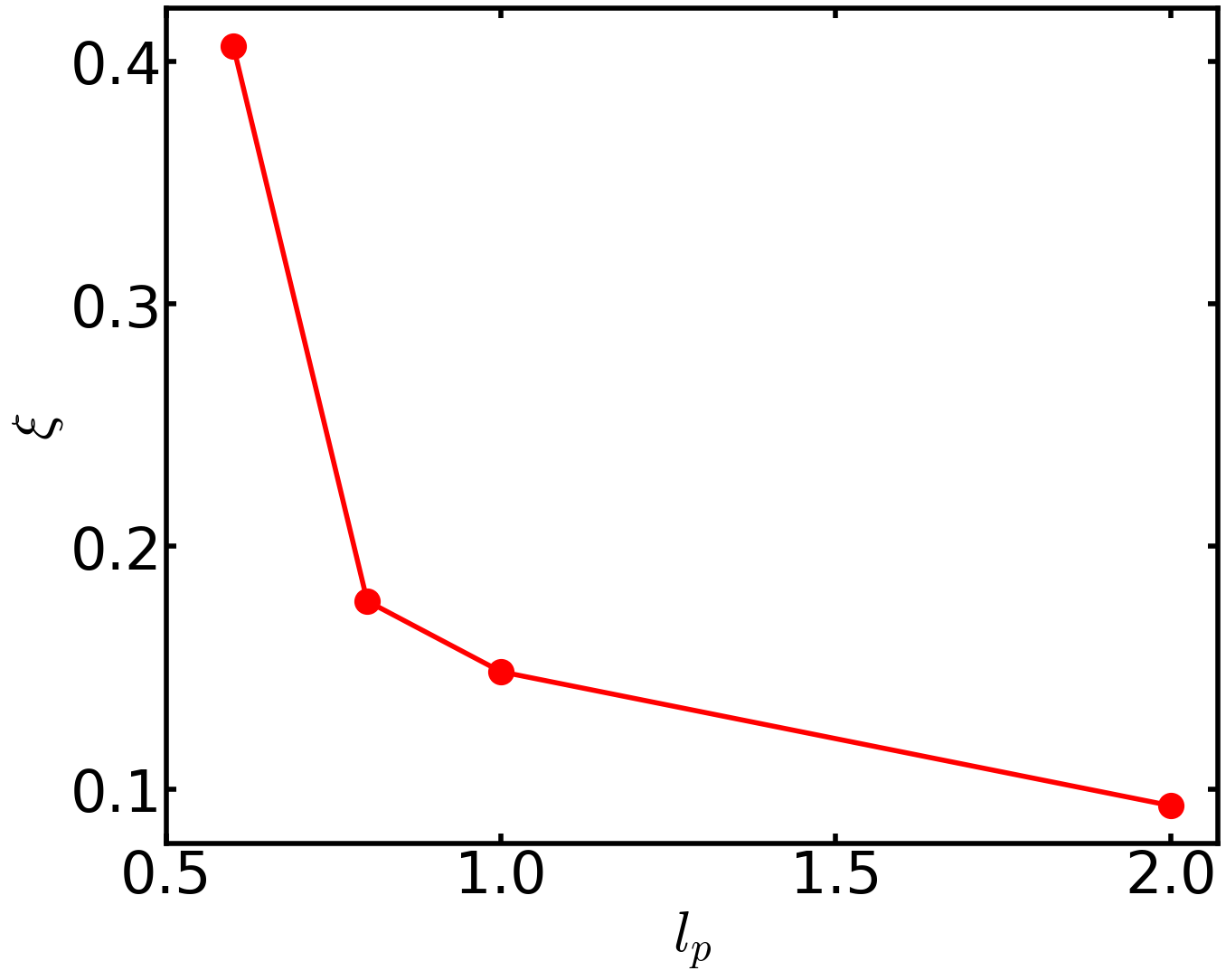}
    \caption{}
    \label{fig:xi-lp}
\end{subfigure}
 \hfill
 \begin{subfigure}{0.3\textwidth}
 \includegraphics[width=\textwidth]{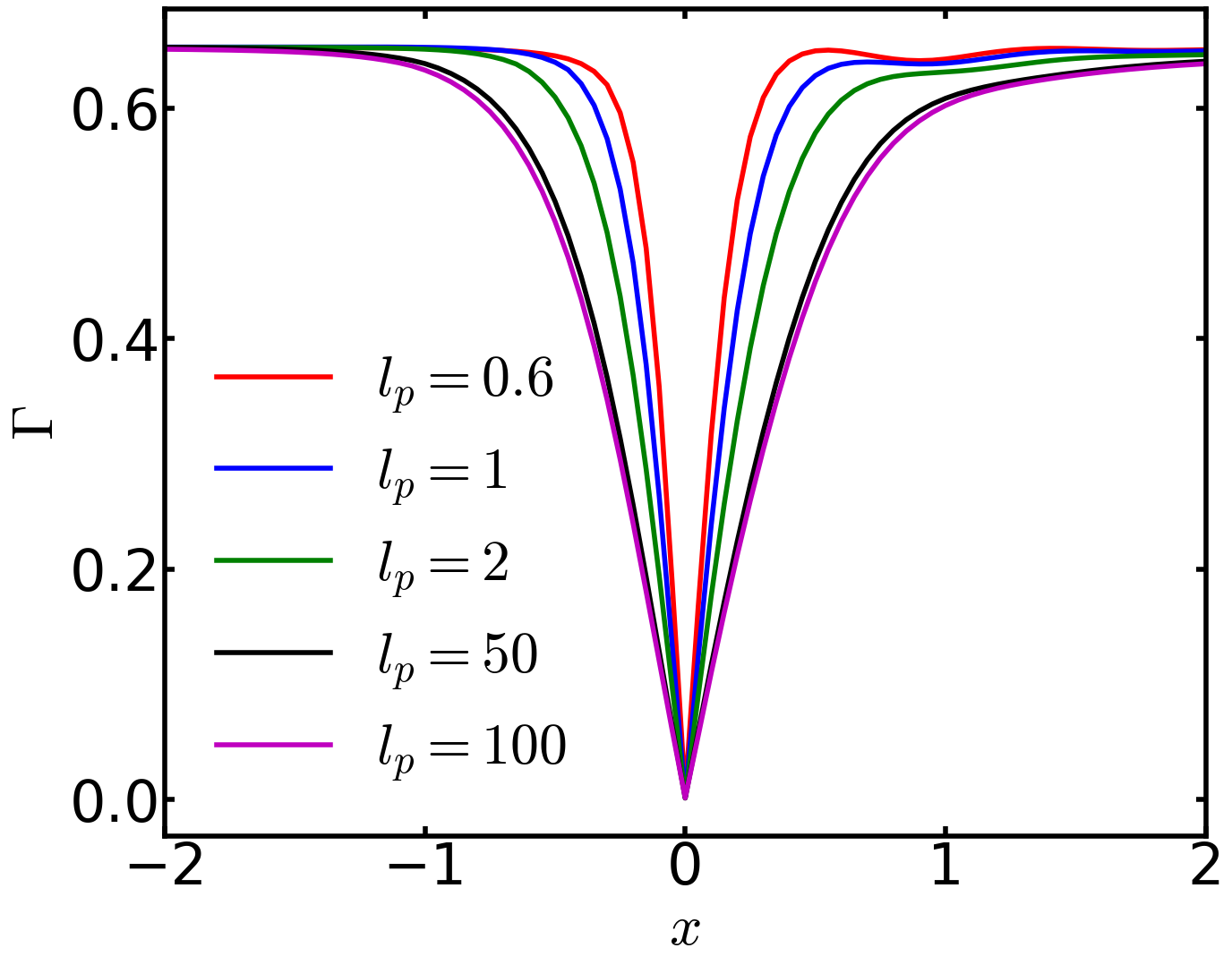}
    \caption{}
    \label{fig:Gamma-lp-fixedS}
\end{subfigure}
 \hfill
\caption{Normalized optical retardance $\Gamma/\Gamma_N$ as a function of position along the line $y=0$ for $-1/2$ defect at varying values of (a)$u_2$ and (b)$l_p$. Charateristic length $\xi$ as a function of (d)$u_2$ and (e)$l_p$. The point for $u_2=24, u_0=15$ appears anomalous due to its coupling to a wide density variation, thus points of $u_0=20$ are added for comparison. (c) and (f): Density and optical retardance along the line $y=0$ for $-1/2$ defect at varying vaules of $l_p$, where $u_0=50$ to make the systems close to the incompressible limit, and $u_2$ are adjusted to make the $S_N\approx0.652$($u_2=22$ for $l_p=0.6$; $u_2=15$ for $l_p=1$; $u_2=9.8$ for $l_p=2$; $u_2=5.46$ for $l_p=50$; $u_2=5.4$ for $l_p=100$).}
\label{fig:cores_2}
\end{figure*}

We finally examine the dependence of the core structure on the interaction coefficients $u_0$ and $u_2$, as well as on chain flexibility. Figure \ref{fig:cores} presents the density profiles (top) and optical retardance (bottom) along $y=0$ for a $q = - 1/2$ defect under varying parameters. For each plot, all parameters are held constant($u_0=15$, $u_2 = 15$, $L_c=1$, $l_p = 1$, and $n/V=1$), except for the specific parameter being examined. We observe that the compressibility parameter $u_0$ does not affect nematic order ($S_N$) in the nematic region. The core radius decreases as the system become less compressible and reaches a plateau in the incompressible limit. However, $S_N$ shows an increase with $u_2$ and $l_p$, as expected from the bulk state study. The density difference between the defect core and the nematic regions decreases as the system becomes less compressible (i.e., with increasing $u_0$). Increases in $u_2$ or $l_p$ lead to greater order ($S_N$) in the nematic region and an increased density difference between the defect core and the nematic regions. However, while an increase in $u_2$ expands the region of density variation, the size of density variation region remains relatively unchanged when $l_p$ is increased. To quantitatively analyze the size of the core from the $\Gamma$ configurations, we replotted Figures \ref{fig:cores}(e) and \ref{fig:cores}(f) in Figures \ref{fig:cores_2}(a) and \ref{fig:cores_2}(b) by normalizing the data using the optical retardance in the nematic region ($\Gamma_N$) and a characteristic length $\xi$, defined such that $\Gamma(x=-\xi, y=0)=\Gamma_N/2 $. It allows $\xi$ to be displayed as a function of $u_2$ and $l_p$, as shown in Figures \ref{fig:cores_2}(d) and \ref{fig:cores_2}(e). The characteristic length $\xi$ decreases with increasing $u_2$ or $l_p$, accompanied by an increase in $\Gamma_N$. The point for $u_2=24, u_0=15$ appears anomalous due to its coupling to a wide density variation. To exclude the effect of $\Gamma_N$ and compressibility on the core size, we chose $u_0=50$ and set $u_2$ in order to make $S_N \approx 0.652$ for various $l_p$. The density and optical retardance profiles along the line $y=0$ are shown in \ref{fig:cores_2}(c)(f). Therefore, for an incompressible systems and fixed $S_N$, the core size increases as the chains become stiffer.

\section{Conclusions}
In this article, we have studied the equilibrium isotropic-nematic phase transition, isotropic-nematic interfaces, and topological defects in the nematic phase of a worm like chain liquid crystal by using Self-Consistent Field Theory. For homogeneous states, the phase transition follows by numerical minimization of the free energy functional, revealing the effects of chain number density and flexibility on the transition. Larger density or stiffer chains are associated with lower Maier-Saupe interaction strength at the transition point. In the case of an isotropic-nematic interface, we have studied the effects of compressibility, interaction strength $u_2$, and chain flexibility on the interface. A larger value of $u_0$ corresponds to a smaller compressibility, leading to a smaller density gap between the two phases. In the canonical ensemble, the value of $u_2$ determines the relative portion of the system occupied by the isotropic and nematic phases. The interfacial width increases as the chain become stiffer, as given by larger persistence length $l_p$. Both  homeotropic and planar alignments have been studied. In contrast to homeotropic alignment, planar alignment displays nonzero biaxiality across the interface and a smaller interfacial width. We have also studied the equilibrium configurations associated with  $\pm 1/2$ disclinations in a thin film geometry (two dimensional variations of the nematic order parameter, but allowing out of plane director orienation). Defects display a region of biaxial order in the core region, and a uniaxial center defined as the point in which $S=P$, the uniaxial and biaxial order parameters coincide. Elastic anisotropy, which is naturally incorporated in the flexible chain model, leads to an anisotropic core in agreement with previous experiments and computation. Finally, we have examined the dependence of the core structure and extent on the interaction coefficients $u_0$ and $u_2$, as well as on chain flexibility. For fixed far field nematic order $S_N$, the core size is seen to increase with $l_p$, and hence to effectively decrease with increasing twist/bend elastic anisotropy.

\section*{Conflicts of interest}

There are no conflicts of interest to declare.

\section*{Acknowledgements} 
This research has been supported by the National Science Foundation under contract DMR-2223707, by the Minnesota Supercomputing Institute of the University of Minnesota, and by the Advanced Cyberinfrastructure Coordination Ecosystem: Services \& Support (ACCESS) program, which is supported by U.S. National Science Foundation grants 2138259, 2138286, 2138307, 2137603, and 2138296.


\appendix

\bibliography{references} 
\bibliographystyle{rsc} 

\end{document}